\documentclass[numberedappendix,twocolumn,tighten]{aastex61}
\usepackage[bookmarks=false]{hyperref}
\usepackage{natbib}
\accepted{April 5, 2018}
\submitjournal{ApJ}

\begin{document}

\title{Spatially Resolved Stellar Kinematics from LEGA-C: Increased Rotational Support in $z{\sim}0.8$ Quiescent Galaxies}

\correspondingauthor{Rachel Bezanson}
\email{ rachel.bezanson@pitt.edu}

\author{Rachel Bezanson}
\affiliation{Department of Physics and Astronomy and PITT PACC, University of Pittsburgh, Pittsburgh, PA, 15260, USA}
\affiliation{Department of Astrophysics, Princeton University, Princeton, NJ 08544, USA}

\author{Arjen van der Wel}
\affiliation{{Sterrenkundig Observatorium, Universiteit Gent, Krijgslaan 281 S9, B-9000 Gent, Belgium}}
\affiliation{Max-Planck Institut f{\"u}r Astronomie, K{\"o}nigstuhl 17, D-69117, Heidelberg, Germany}

\author{Camilla Pacifici}
\affiliation{Space Telescope Science Institute, 3700 San Martin Drive, Baltimore, MD 21218, USA}

\author{Kai Noeske}
\affiliation{Max-Planck Institut f{\"u}r Astronomie, K{\"o}nigstuhl 17, D-69117, Heidelberg, Germany}

\author{Ivana Bari\v{s}i\'c}
\affiliation{Max-Planck Institut f{\"u}r Astronomie, K{\"o}nigstuhl 17, D-69117, Heidelberg, Germany}

\author{Eric F. Bell}
\affiliation{Department of Astronomy, University of Michigan, 1085 South University Ave., Ann Arbor, MI 48109, USA}

\author{Gabriel B. Brammer}
\affiliation{Space Telescope Science Institute, 3700 San Martin Drive, Baltimore, MD 21218, USA}

\author{Joao Calhau}
\affiliation{Department of Physics, Lancaster University, Lancaster LA1 4YB, UK}

\author{Priscilla Chauke}
\affiliation{Max-Planck Institut f{\"u}r Astronomie, K{\"o}nigstuhl 17, D-69117, Heidelberg, Germany}

\author{Pieter van Dokkum}
\affiliation{Astronomy Department, Yale University, New Haven, CT 06511, USA}

\author{Marijn Franx}
\affiliation{Leiden Observatory, Leiden University, P.O.Box 9513, NL-2300 AA Leiden, The Netherlands}

\author{Anna Gallazzi}
\affiliation{INAF-Osservatorio Astrofisico di Arcetri, Largo Enrico Fermi 5, I-50125 Firenze, Italy}

\author{Josha van Houdt}
\affiliation{Max-Planck Institut f{\"u}r Astronomie, K{\"o}nigstuhl 17, D-69117, Heidelberg, Germany}

\author{Ivo Labb\'e}
\affiliation{Leiden Observatory, Leiden University, P.O.Box 9513, NL-2300 AA Leiden, The Netherlands}

\author{Michael V. Maseda}
\affiliation{Leiden Observatory, Leiden University, P.O.Box 9513, NL-2300 AA Leiden, The Netherlands}

\author{Juan Carlos Mu\~nos-Mateos}
\affiliation{European Southern Observatory, Alonso de Crdova 3107, Casilla 19001, Vitacura, Santiago, Chile}

\author{Adam Muzzin}
\affiliation{Department of Physics and Astronomy, York University, 4700 Keele St., Toronto, Ontario, Canada, MJ3 1P3}

\author{Jesse van de Sande}
\affiliation{Sydney Institute for Astronomy, School of Physics, A28, The University of Sydney, NSW, 2006, Australia}

\author{David Sobral}
\affiliation{Department of Physics, Lancaster University, Lancaster LA1 4YB, UK}
\affiliation{Leiden Observatory, Leiden University, P.O.Box 9513, NL-2300 AA Leiden, The Netherlands}

\author{Caroline Straatman}
\affiliation{{Sterrenkundig Observatorium, Universiteit Gent, Krijgslaan 281 S9, B-9000 Gent, Belgium}}

\author{Po-Feng Wu}
\affiliation{Max-Planck Institut f{\"u}r Astronomie, K{\"o}nigstuhl 17, D-69117, Heidelberg, Germany}

\shortauthors{Bezanson et al.}
\shorttitle{LEGA-C: Increased Rotational Support in Quiescent Galaxies at $z\sim0.8$}

\begin{abstract} 
We present stellar rotation curves and velocity dispersion profiles for 104 quiescent galaxies at $z=0.6-1$ from the Large Early Galaxy Astrophysics Census (LEGA-C) spectroscopic survey. Rotation is typically probed across 10-20kpc, or to an average of 2.7$R_e$. Combined with central stellar velocity dispersions ($\sigma_0$) this provides the first determination of the dynamical state of a sample selected by a lack of star formation activity at large lookback time. The most massive galaxies ($M_{\star}>2\times10^{11}\,M_{\odot}$) generally show no or little rotation measured at 5kpc ($|V_5|/\sigma_0<0.2$ in 8 of 10 cases), while ${\sim}64\%$ of less massive galaxies show significant rotation. This is reminiscent of local fast- and slow-rotating ellipticals and implies that low- and high-redshift quiescent galaxies have qualitatively similar dynamical structures. We compare $|V_5|/\sigma_0$ distributions at $z\sim0.8$ and the present day by re-binning and smoothing the kinematic maps of 91 low-redshift quiescent galaxies from the CALIFA survey and find evidence for a decrease in rotational support since $z\sim1$. This result is especially strong when galaxies are compared at fixed velocity dispersion; if velocity dispersion does not evolve for individual galaxies then the rotational velocity at 5kpc was an average of ${94\pm22\%}$ higher in $z\sim0.8$ quiescent galaxies than today. Considering that the number of quiescent galaxies grows with time and that new additions to the population descend from rotationally-supported star-forming galaxies, our results imply that quiescent galaxies must lose angular momentum between $z\sim1$ and the present, presumably through dissipationless merging, and/or that the mechanism that transforms star-forming galaxies also reduces their rotational support.
\end{abstract}

\section{Introduction}
Massive galaxies exhibit significant angular momentum at $z\sim2$ \citep[e.g.,][]{forster:09, forster:11, tacconi:13, wel:14_shapes, dokkum:15, wisnioski:15, belli:17, straatman:17}. This statement is in contrast with massive galaxies in the local Universe, which are predominantly elliptical, where even so-called ``fast-rotators'' exhibit significant dispersion support \citep[e.g.,][]{emsellem:07,emsellem:11}. This discrepancy necessitates an evolution in rotational support through cosmic time, however how and why this change occurred remains up for debate. One possibility is that the quenching process itself destroys organized rotation and/or the destruction of organized rotation is effectively what quenches galaxies \citep[e.g.,][]{hopkins:08,martig:09}. Alternatively the evolution could be more gradual, owing to subsequent minor or major merging \citep[e.g.,][]{naab:09,hilz:13,naab:14}. We have evidence that the latter must play some role from the size evolution of quiescent, or ``non-star-forming'', galaxies, which grow significantly in size on average through cosmic time \citep[e.g.,][and references within]{daddi:05,toft:07,trujillo:07,dokkumnic:08,wel:08, newman:12, wel:14}. This  growth is most likely due to dissipationless minor merging \citep[e.g.,][]{bezanson:09,hopkins:09cores,naab:09,dokkum:10,oser:11} which would diminish net angular momentum.  For a given population of quiescent galaxies the rate of stellar rotation should then decrease with cosmic time, implying that high-redshift quiescent galaxies would show
more rotation than their present-day counterparts. 

A complicating factor is that the high- and low-redshift populations cannot be directly compared due to the increase in the number of quiescent galaxies with cosmic time, as galaxies cease to form stars, or ``quench''.  This effect, often called \emph{progenitor bias}, has been investigated thoroughly as a potential driver of the empirical size evolution of quiescent galaxies as new and more extended additions to the red sequence would drive evolution in the average size-mass relations \citep[e.g.,][]{wel:09, valentinuzzi:10a, valentinuzzi:10b, poggianti:13, carollo:13,lilly:16,fagioli:16, williams:16}. Although there is evidence for some evolution in the velocity dispersions of star-forming galaxies through cosmic time, these galaxies have been shown to be primarily rotating disks since at least $z\sim2$ \citep[e.g.,][]{wisnioski:15,simons:17}. Without any structural transformation, these new additions would also represent an influx of still-rotating quiescent galaxies as they have had less time since quenching to lose their angular momentum than their older counterparts.  Therefore, the fraction of rotating galaxies in the quiescent population may increase over cosmic time.  One further level of complexity in this picture is that quenching of star formation may coincide with a change
in dynamical structure, as suggested by the smaller relative sizes of post-starburst galaxies that constitute the newest additions to the high-redshift quiescent population \citep{whitaker:12a,yano:16}. 

So far, evolution in the shape distribution of quiescent galaxies has provided one of the strongest constraints on the evolution of angular momentum among the population of quiescent galaxies.  The emerging picture is that oblate, flat shapes are more common among high-redshift quiescent galaxies than in the present-day universe \citep{wel:11,chevance:12,chang:13}.  The (projected) shape of
a galaxy is obviously only a crude proxy of dynamical structure, and even for large samples the necessary assumption was made that the population of high-redshift quiescent galaxies was composed of galaxies with the same intrinsic shapes as today's galaxies: oblate disks and triaxial spheroids.  The relative numbers of both types were then inferred to change with redshift \citep{chang:13}.

However, it is not self-evident that galaxy structures are the same at
different cosmic epochs and the correspondence between global shape and
kinematic properties may well evolve.  Therefore, it is essential to obtain
spatially resolved kinematics of high-redshift quiescent galaxies, which must be measured from stellar absorption features.
Currently, such direct evidence comes from small samples without uniform or necessarily representative selections. These include two examples of
strongly lensed galaxies at $z\sim 2$ \citep{newman:15,toft:17} and samples of 25 $z\sim0.5$ cluster \citep{marel:07,moran:07} and $z\sim 1$ field galaxies \citep{wel:08_rot}.  The latter samples were selected on the basis of visual morphology, that is, a visual
determination of the absence of a disk-like structure, preventing a rigorous analysis of the evolution of rotation among quiescent galaxies at different epochs. Finally, \citet{belli:17} found indirect evidence of evolution in the rotational support of quiescent galaxies from dynamical masses. In this paper we present a much larger sample of ${\sim}100$ galaxies at $z\sim 0.8$ selected by their lack of star-formation activity and with high-quality stellar rotation curves from the Large Early Galaxy Astrophysics Census (LEGA-C) survey.

This paper begins in \S \ref{sect:legac} with a brief description of the LEGA-C survey and the extraction of spatially resolved stellar kinematics. In \S \ref{sect:results} we investigate the empirically derived rotational support of massive quiescent galaxies at $z\sim0.8$ and the trends of that rotation with galaxy properties derived from imaging data. In \S \ref{sect:califa} we use stellar kinematics derived from the CALIFA DR3 dataset to assess the effects of seeing on the LEGA-C observations and study the redshift-evolution of the rotational support of massive quiescent galaxies. Finally in \S \ref{sect:discussion}, we conclude with a discussion of these results in the context of models of galaxy evolution and other observational and theoretical studies. Throughout this paper we assume a standard concordance cosmology ($H_0 = 70 \mathrm{km\,s^{-1}\,Mpc^{-1}}$, $\Omega_M = 0.3$, and $\Omega_{\Lambda}=0.7$).

\section{LEGA-C Data and Stellar Kinematics}\label{sect:legac}

\subsection{The LEGA-C Spectroscopic Survey}

\begin{figure*}[t]
    \centering
    \includegraphics[width=0.9\textwidth]{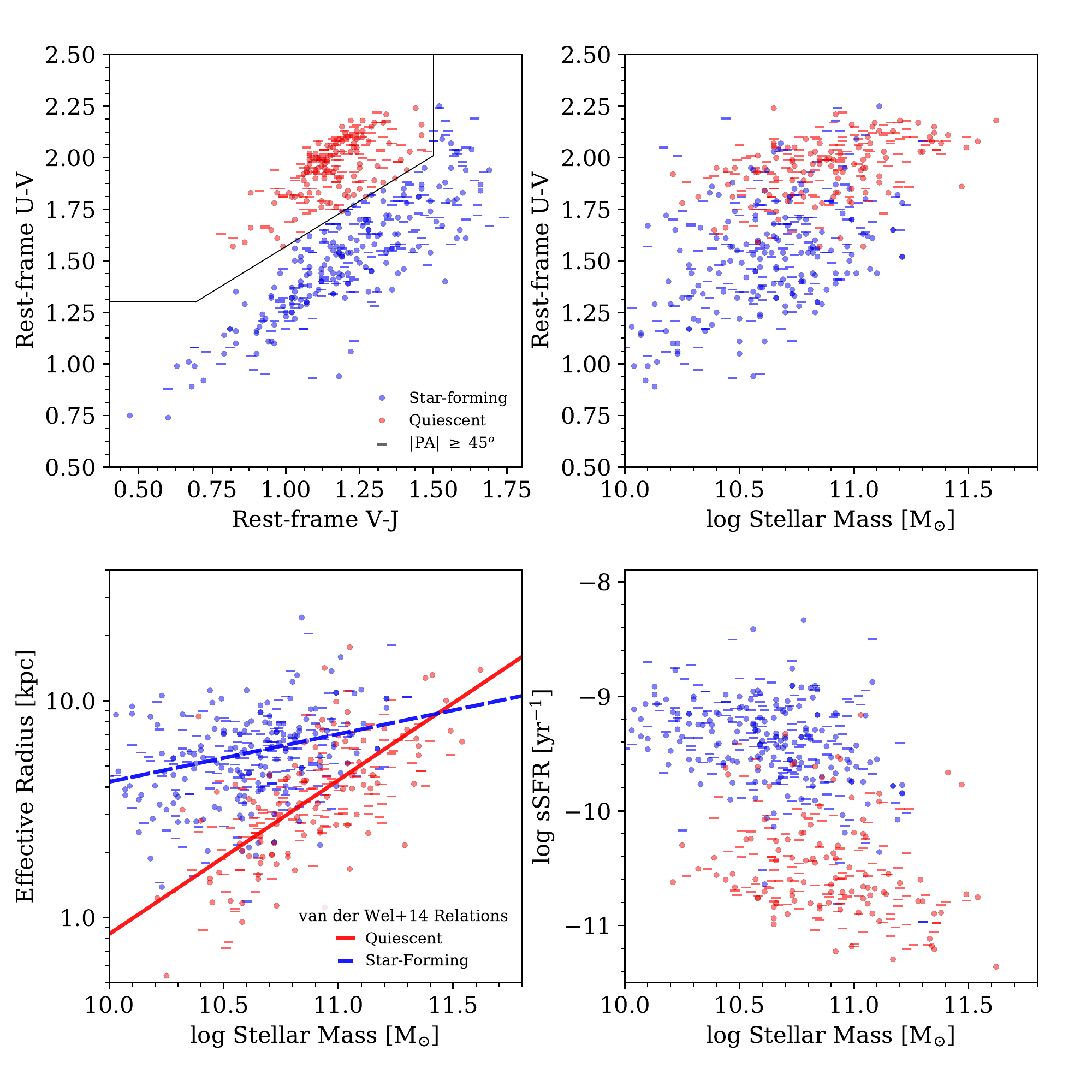}
    \caption{Properties of the complete LEGA-C year one dataset. Symbol colors differentiate between star-forming (blue) and quiescent (red) galaxies as determined by U-V and V-J rest-frame colors and cuts from \citet{muzzin:13} (upper left panel). Misaligned galaxies ($|PA|\geq45^o$) are excluded from this study and are indicated by horizontal lines. Star-forming and quiescent galaxies in the LEGA-C sample have different distributions in color (upper right panel), physical size (bottom left panel), and sSFR (bottom right panel); for this study we focus on the kinematics of the quiescent population.}
    \label{fig:sample}
\end{figure*}

The spectroscopic data included in this analysis are drawn from the first year data release of the Large Early Galaxy Astrophysics Census (LEGA-C) survey \citep{wel:16}. This project is a 128-night ESO Public Spectroscopic survey of massive galaxies at $0.6<z<1.0$ in the COSMOS field using VIMOS on the VLT. The LEGA-C Survey primary sample of $\sim\,3000$~galaxies is selected with a photometric or spectroscopic redshift-dependent K-magnitude limit ($K = 20.7-7.5\times\log((1+z)/1.8)$), corresponding to a representative sampling of galaxy colors down to $\log M_{\star}/M_{\odot} \gtrsim 10.4$.  The defining, unique aspect of the LEGA-C spectra is the deep 20-hour long integration at a resolution of $R=2500$ in the wavelength range ${\sim}6300-8800\mathrm{\AA}$. The first year dataset consists of 7 masks of roughly 130 galaxies in each mask with slits that are oriented in the N-S direction. The combined data yield the extremely high signal-to-noise ${S/N\sim20\mathrm{\AA}^{-1}}$ in the continuum. The data reduction procedure is described by \citet{wel:16}. Two-dimensional and extracted one-dimensional reduced spectra are publicly available via the ESO Science Archive Facility.

\subsection{Photometry: Stellar Populations and Structures}
Additional ancillary data are available for the LEGA-C sample in the COSMOS field. Targeted galaxies are selected from the UltraVISTA version DR1 4.1 K-selected catalogs \citep{muzzin:13ultravista}. Rest-frame colors are calculated from the UltraVISTA photometry \citep{mccracken:12,muzzin:13ultravista} using the EAZY \citep{eazy} code and fixing redshifts to the LEGA-C spectroscopic redshifts. Stellar population properties, most notably stellar masses, are determined from the UltraVISTA photometry  using the FAST code \citep{kriek:09} and using \citet{bc:03} stellar population libraries, adopting a \citet{chabrier:03} Initial Mass Function (IMF), \citet{calzetti:00} dust extinction, and exponentially declining star-formation rates. Although formal uncertainties on stellar masses are relatively low, systematics likely dominate and we adopt an uncertainty of $0.2$\,dex following \citet{muzzin:09}. Star-formation rates are estimated from the UV and IR (24$\mu m$ from Spitzer-MIPS) luminosities, following \citet{whitaker:12b}. Morphological information is derived for all galaxies from COSMOS Hubble Space Telescope (HST) ACS F814W imaging \citep{cosmosacs, massey:10}, which is well matched to the rest-frame optical at this redshift. Best-fit S\'{e}rsic parameters, and uncertainties are derived for all LEGA-C galaxies using GALFIT \citep{galfit} and GALAPAGOS \citep{galapagos} following the procedures outlined in \citet{wel:12} and \citet{wel:16}. The quoted measurement uncertainties of structural parameters do not include a number of systematic uncertainties and specifically do not account for covariance of parameters, which could dominate e.g. for S\'ersic parameters. For visual presentation, we match the LEGA-C catalog to imaging from the first public data release of the Hyper-Suprime Cam Subaru Strategic Program (HSC-SSP), which includes deep \textit{grizy} imaging in the COSMOS field \citep{aihara:17}.

\begin{figure*}[!t]
    \centering
    \includegraphics[width=0.95\textwidth]{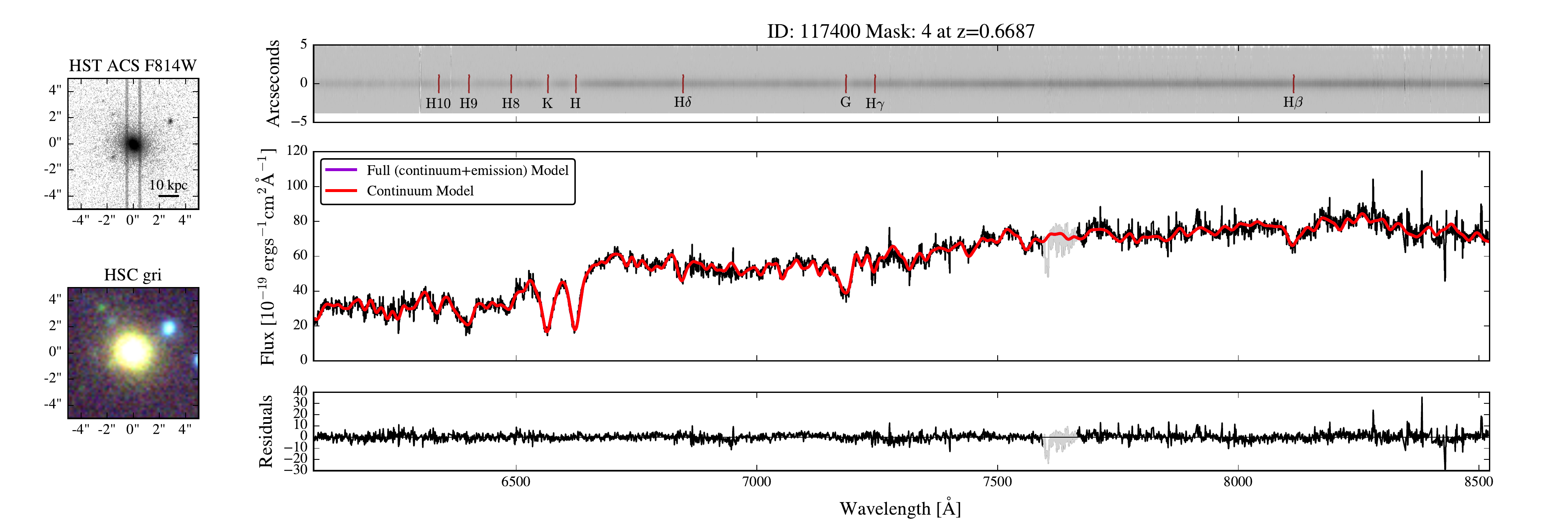}
    \includegraphics[width=0.95\textwidth]{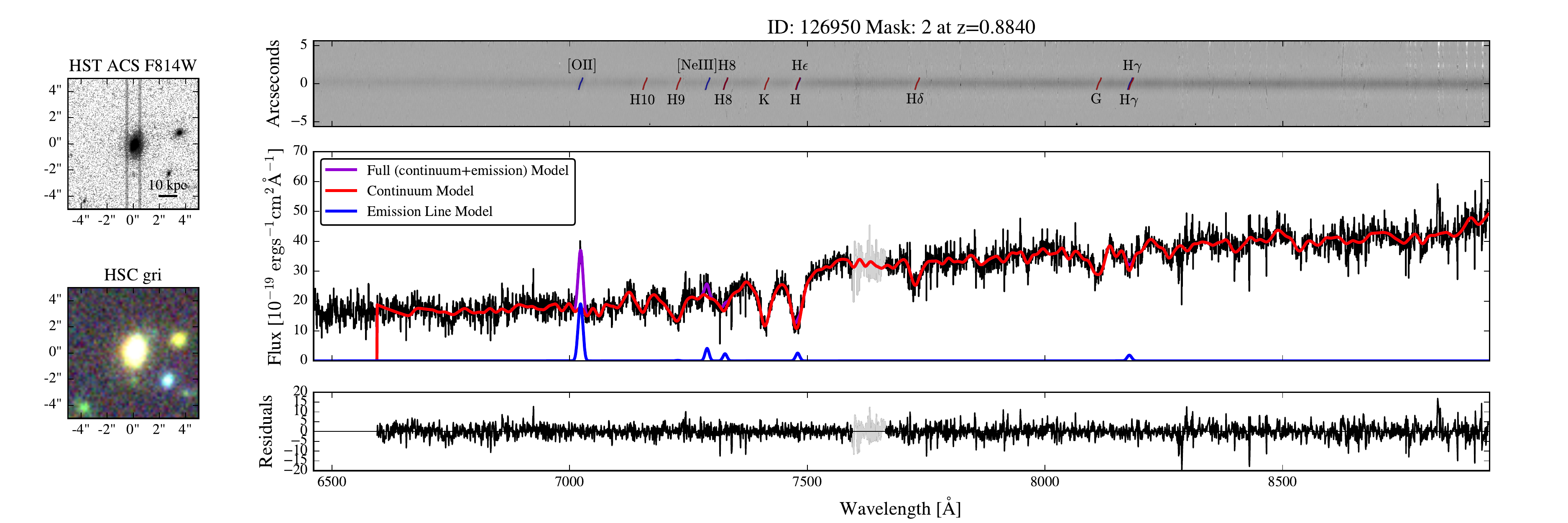}
    \includegraphics[width=0.95\textwidth]{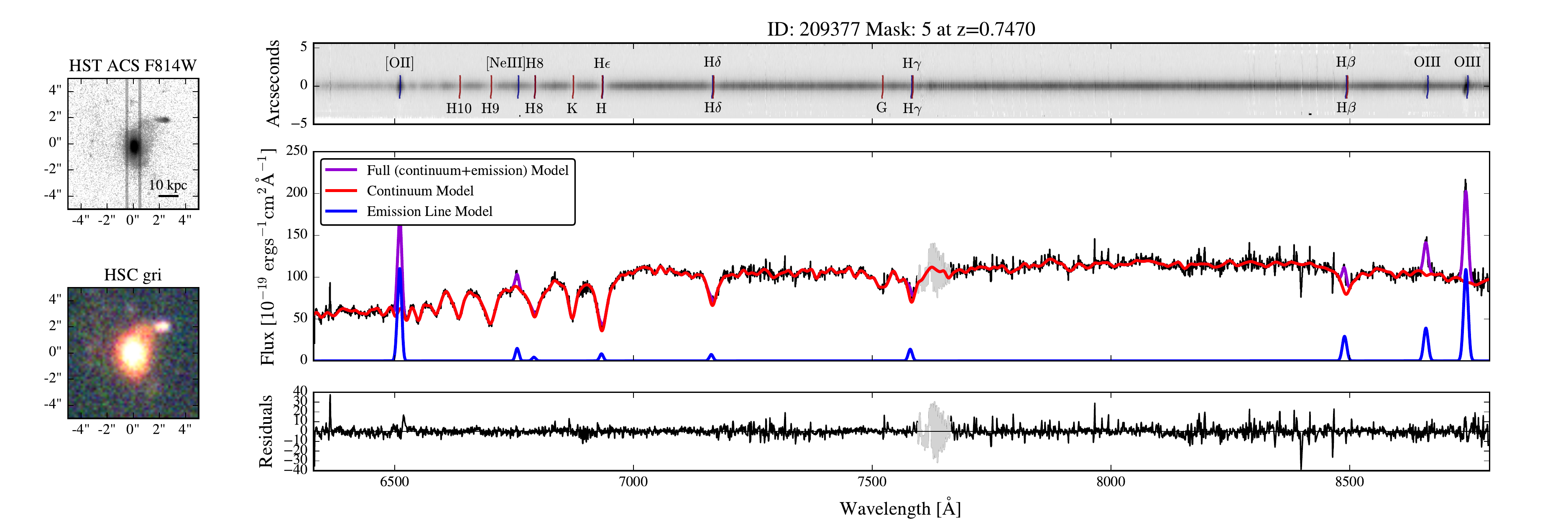}
    \caption{Images and spectra of three example quiescent galaxies from the LEGA-C sample, selected to span a range in emission line flux for demonstration of fitting; most galaxies in the sample do not exhibit significant emission. Images from HST ACS COSMOS mosaics and gri color images from the Hyper-Suprime Cam SSP public data release. The position and width of the LEGA-C slit as well as the physical scale are indicated on the HST image. The top panel in each row shows the 2D LEGA-C spectrum, with the location of spectral absorption and emission features, including the measured rotation, indicated with blue and red lines. Emission line features are labeled above the galaxy spectrum and continuum features are indicated below. One-dimensional optimally extracted spectra are included in the middle panel to demonstrate the continuum plus emission-line modeling.  Best-fit continuum models are indicated by red lines, emission lines, where detected, are indicated by blue lines, and the combined model by purple lines. Residuals from the 1-D fit are included in the bottom panel. In this work, this procedure is repeated separately on all rows with sufficient S/N in the 2D spectra.}
    \label{fig:spectra}
\end{figure*}

Stellar population and structural properties for the year one LEGA-C massive galaxies are shown in Figure \ref{fig:sample}. Symbol color indicates whether galaxies are categorized as quiescent (red) or star-forming (blue) based on their rest-frame $U-V$ and $V-J$ colors (upper left panel), adopting the \citet{muzzin:13} color cuts, which are specifically defined for the UltraVISTA photometric catalogs used in determining rest-frame colors. We note that although this selection does a very good job of identifying galaxies with quiescent stellar populations, there are a subset of galaxies in the current sample with clearly detected emission lines (see Figure \ref{fig:spectra} for examples). 

Horizontal lines indicate galaxies included in the full sample for which the semi-major axis is significantly inclined with respect to the VIMOS slits ($|PA|\geq45^o$). These objects are not considered in this paper, as the mismatch between the kinematic axis and the slit will prevent us from tracing stellar rotation in a straightforward manner \citep[e.g.,][]{weiner:06,straatman:17}. For this paper we focus on major axis kinematics in the 104 quiescent galaxies for which the major axes are aligned to within $|PA|<45^o$ of the N-S slits (circles in Figure \ref{fig:sample}).  We note that the quoted position angles are photometric and the kinematic axes can also be misaligned with the photometry \cite[e.g.,][]{franx:89,emsellem:07}. \citet{emsellem:07} demonstrated that for fast rotators this effect is minimal ($\lesssim 10\%$), but kinematic and photometric position angles can be significantly misaligned, by up to ${\sim}50\%$ in the SAURON sample.  However \citet{krajnovic:11} showed that for 90\% of galaxies in the ATLAS3D sample, the kinematic misalignment will be $\leq15^o$.

The top right panel of Figure \ref{fig:sample} shows the rest-frame $U-V$ colors of the two populations as a function of stellar mass. The bottom left panel shows the effective radius along the semi-major axis versus stellar mass, with the solid, red diagonal line indicating the \citet{wel:14} size-mass relation for quiescent galaxies and dashed blue line for star-forming galaxies. Finally, the bottom right panel shows the specific star formation rate (sSFR) versus stellar mass. We note here that SFRs determined for quiescent galaxies are notably uncertain and as 24 micron flux may be undetected or ambiguous, these sSFRs are likely to be upper limits for our sample of galaxies. The quiescent and star-forming galaxies in the LEGA-C sample exhibit different distributions in all four phase spaces, although the populations overlap slightly in all but the U-V and V-J colors, which  are used to initially differentiate between them.  

\subsection{Spatially Resolved Stellar Kinematics}
\begin{figure*}[]
    \centering
    \includegraphics[width=0.99\textwidth]{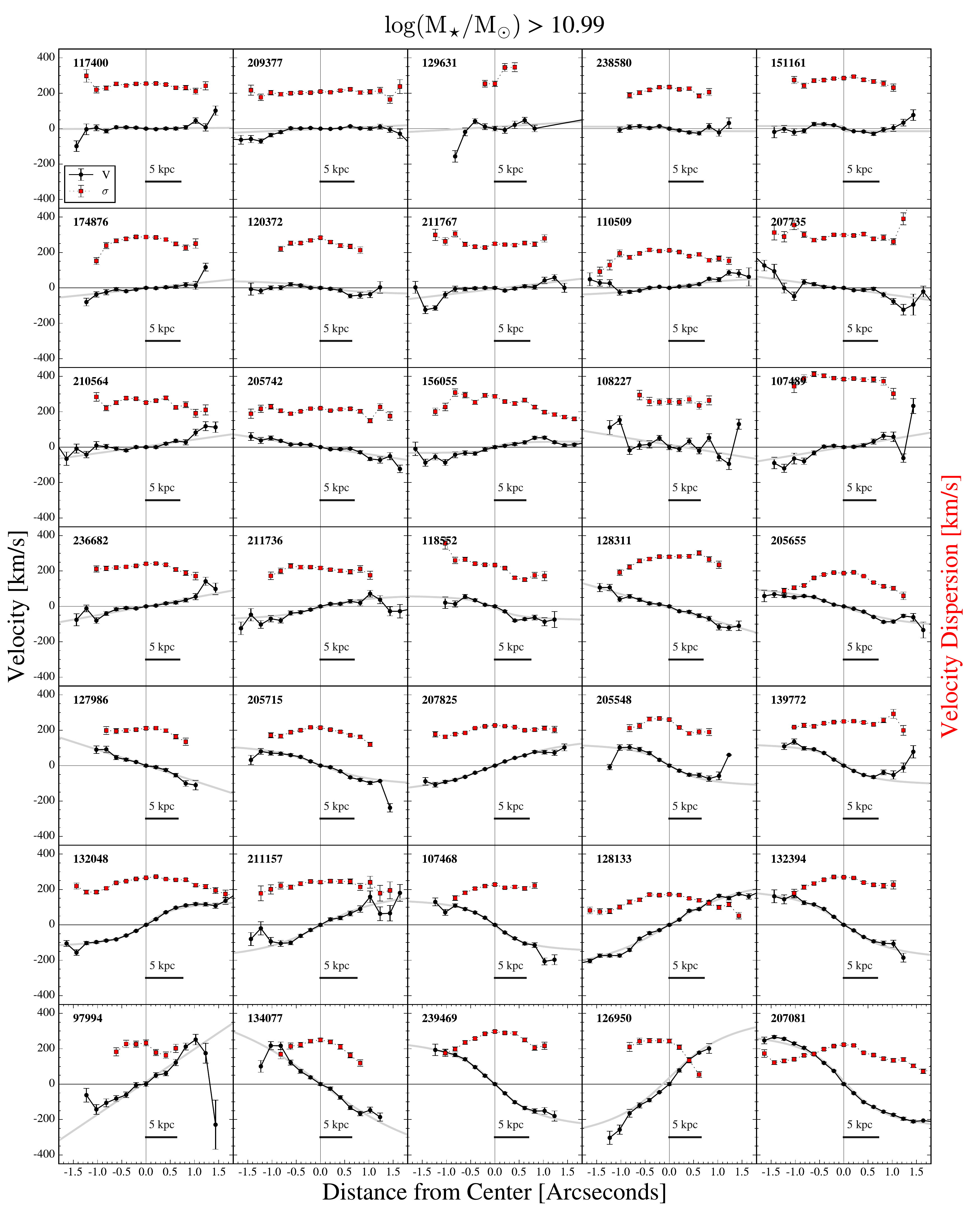}
    \caption{Stellar rotation curves (black) and velocity dispersion profiles (red) for the 35 highest mass ($\log M_*/M_{\odot} > 11$) quiescent galaxies, ordered by increasing $V_5$. The rotational velocity is defined as the velocity of the best-fitting arctangent function (indicated by the gray solid lines) at a radius of 5 kpc (indicated by the black bars) from the central pixel.}
\end{figure*}

\begin{figure*}[]
    \centering
    \addtocounter{figure}{-1}
    \includegraphics[width=0.99\textwidth]{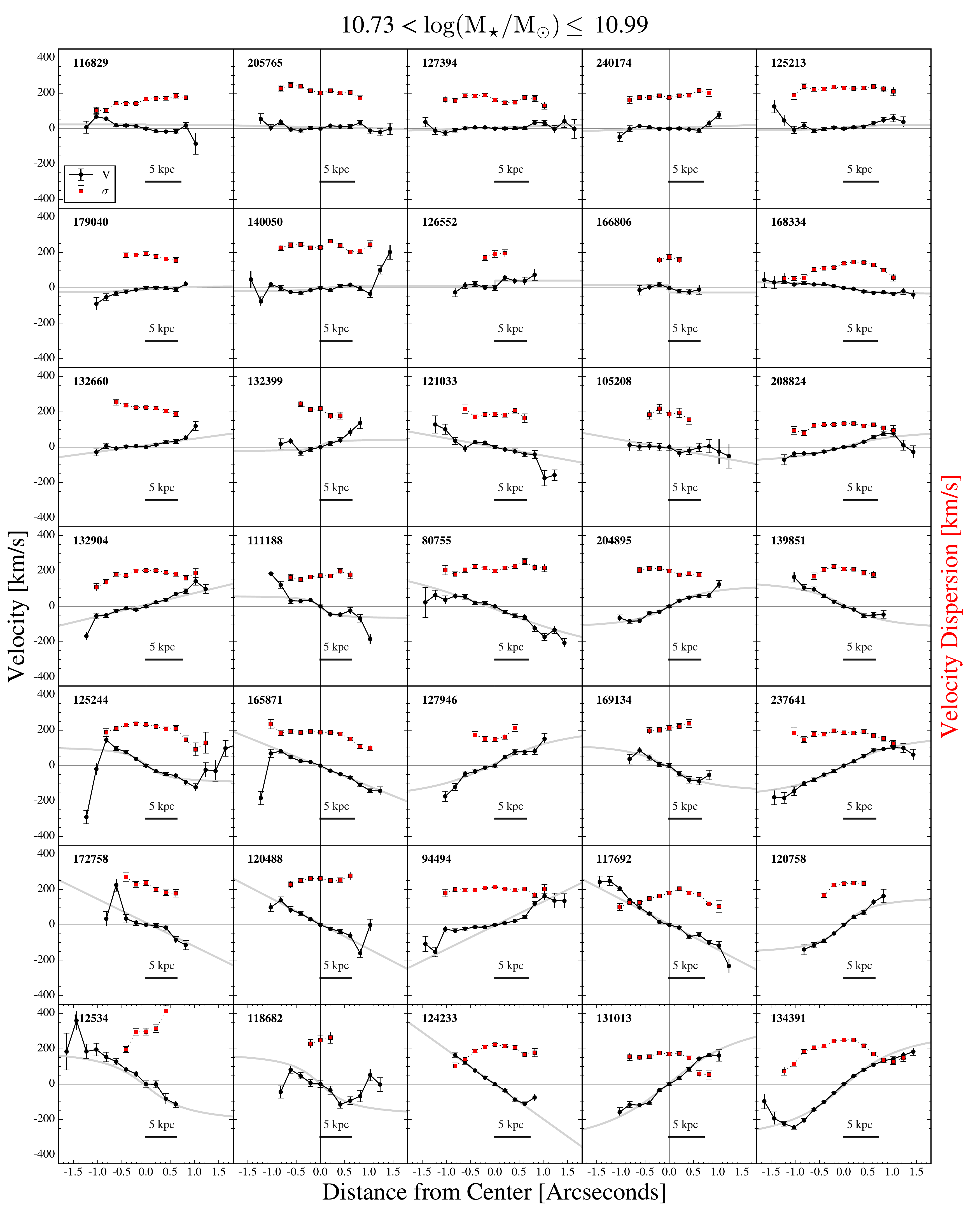}
    \caption{(Continued) Stellar rotation curves (black) and velocity dispersion profiles (red) for the 35 intermediate mass ($10.7 < \log M_*/M_{\odot} \leq 11$) quiescent galaxies.}
\end{figure*}

\begin{figure*}[]
    \centering
    \addtocounter{figure}{-1}
    \includegraphics[width=0.99\textwidth]{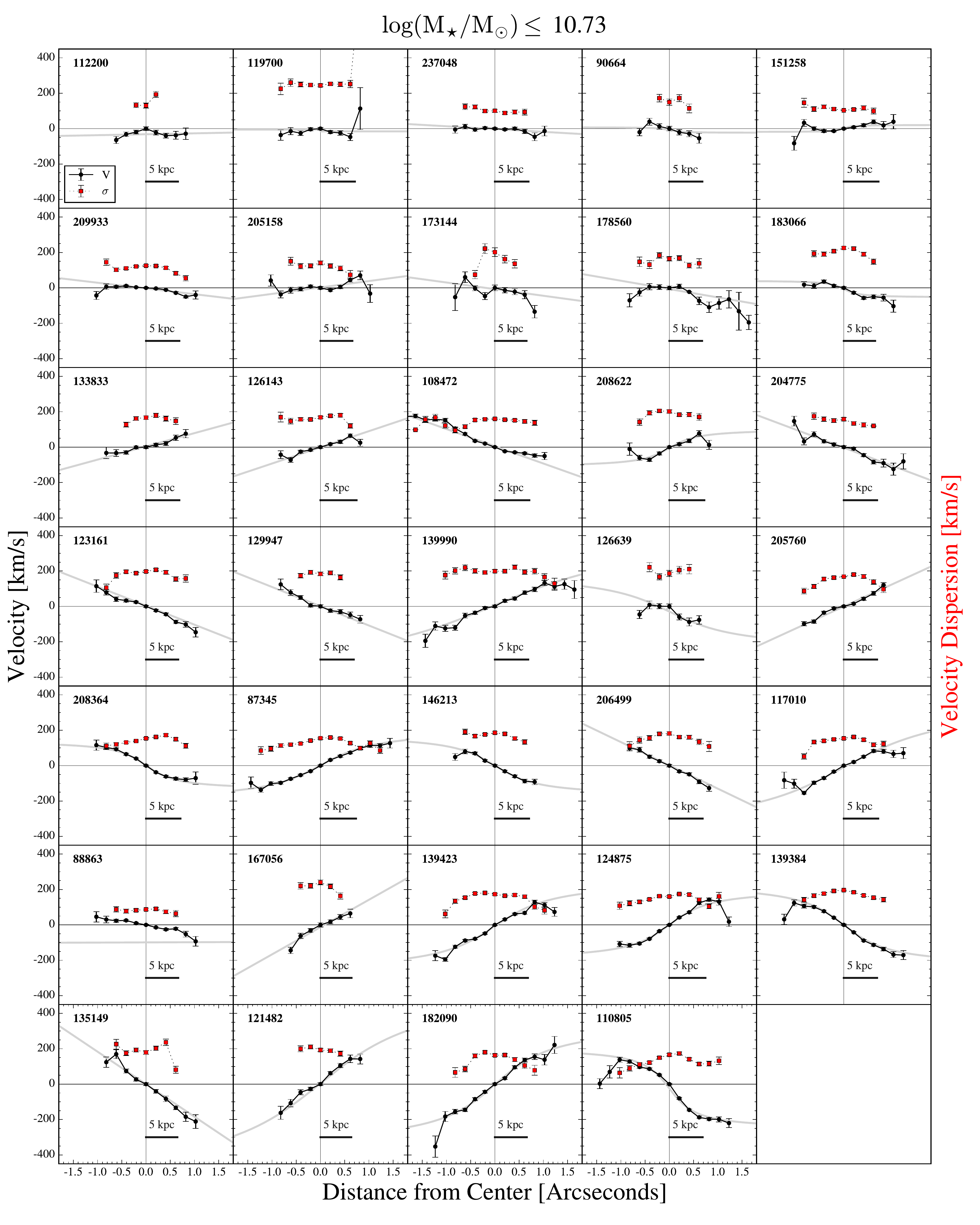}
    \caption{ (Continued) Stellar rotation curves (black) and velocity dispersion profiles (red) for the lowest mass ($\log M_*/M_{\odot} < 10.7$) sample of quiescent galaxies in LEGA-C.}
    \label{fig:rot}
\end{figure*}

We measure the stellar and gas phase line-of-sight kinematics for each galaxy using the Penalized Pixel-Fitting (pPXF) method \citep{cappellari:04} with the updated Python routines \citep{cappellari:17}. For each two-dimensional LEGA-C spectrum, each row with median $S/N>2$ per pixel is fit with two template sets that are allowed to independently shift and broaden: stellar population templates to fit the continuum and a collection of possible emission lines to fit the ionized gas emission. The stellar template is a linear, optimal non-negative combination of \citet{vazdekis:99} single stellar population (SSP) models, which are based on the Medium resolution INT Library of Empirical Spectra (MILES) \citep{miles} empirical stellar spectra combined using \citet{padova:00} isochrones. We extend the rotation curve measurements in the outer rows (with $S/N<2$) by fixing the velocity dispersion to $\sigma=150\,\mathrm{km\,s^{-1}}$ and allowing the normalization and velocity offsets to vary for both stellar and gas templates and find stable results to $S/N\gtrsim1.2$ per pixel. This yields line-of-sight velocity measures out to an average of 8.8 kpc or $2.7R_e$. { We verify that fixing the velocity dispersion to the nearest measured value does not significantly alter the measured rotational velocities, on average leading to a 2\% ($\Delta V=1.2\,\mathrm{km\,s^{-1}}$) offset, which is well within the measurement uncertainty.}

Emission line templates are treated as a single kinematic component, but the normalization of each line (Balmer lines: H10, H9, H8, H$\epsilon$, H$\delta$, H$\gamma$, H$\beta$, H$\alpha$; [NeV], [NeVI], [NeIII], and [OII] and [OIII] doublets) is a free parameter in the fit. The optimally extracted 1-D spectra and best-fit models are shown in Figure \ref{fig:spectra} for three galaxies with increasing emission line components. These galaxies are representative (e.g. in S/N), but are selected to demonstrate the necessity of including emission lines in the kinematic fits and the ability of the data to identify emission lines as they fill in broader absorption features. The majority of galaxies in the sample do not have detected emission lines. Images of each galaxy are shown on the left from the COSMOS HST v2.0 ACS Mosaics (top) \citep{cosmosacs} and \textit{gri} composite color images from the Hyper-Suprime Cam Subaru Strategic Program \citep[HSC-SSP][]{aihara:17}. The two-dimensional spectrum is included in the top panel for each galaxy, with the best-fit rotation curve derived from stellar kinematics at the position of a number of strong absorption (red lines) and emission (blue lines) features overplotted. The middle panel shows the one-dimensional optimally extracted spectrum for each galaxy with the best-fitting continuum model in red and for the second and third galaxies the emission line and total models in blue and purple. The bottom panel in each row shows the residuals from the fit, which are minimal and in most cases uncorrelated.

\begin{deluxetable*}{ccccccccccc}[]\label{tbl:props}
\tabletypesize{\scriptsize}
\tablecaption{Properties of Quiescent Galaxies in the LEGA-C Sample}
\tablehead{
\colhead{ID}  &  \colhead{$\mathrm{z_{spec}}$}  &  \colhead{$\mathrm{\log (\frac{M_{\star}}{M_{\odot}})}$}  &  \colhead{$R_e$}  &  \colhead{b/a}  &  \colhead{PA}  &  \colhead{n}  &  \colhead{$V_5$}  &  \colhead{$V_{Re}$}  &  \colhead{$V_{max}$}  &  \colhead{$\sigma_0$} \\ 
 & & & \colhead{[arcsec]} & & \colhead{[degrees]} & & \colhead{[km/s]} & \colhead{[km/s]} & \colhead{[km/s]} & \colhead{[km/s]}
 \\ 
 \colhead{(1)} & \colhead{(2)} & \colhead{(3)} & \colhead{(4)} & \colhead{(5)} & \colhead{(6)} & \colhead{(7)} & \colhead{(8)} & \colhead{(9)} & \colhead{(10)} & \colhead{(11)}}
\startdata
80755 & 0.7325 & 10.9 & 1.05$\pm$0.02 & 0.51$\pm$0.00 & 27.54$\pm$0.45 & 3.90$\pm$0.08 & -68.6$\pm$4.6 & -61.9$\pm$4.4 & -116.4$\pm$9.7 & 197.0$\pm$8.4\\ 
87345 & 0.6226 & 10.7 & 0.67$\pm$0.00 & 0.54$\pm$0.00 & -35.46$\pm$0.26 & 3.13$\pm$0.04 & 89.7$\pm$2.1 & 51.9$\pm$2.2 & 138.7$\pm$2.1 & 157.5$\pm$5.2\\ 
88863 & 0.8124 & 10.4 & 0.20$\pm$0.00 & 0.91$\pm$0.01 & -32.43$\pm$0.61 & 2.62$\pm$0.06 & -99.0$\pm$-99.0 & -99.0$\pm$-99.0 & -99.0$\pm$-99.0 & 96.8$\pm$7.1\\ 
90664 & 0.7480 & 10.2 & 0.07$\pm$0.00 & 0.50$\pm$0.01 & -5.14$\pm$1.50 & 4.66$\pm$0.17 & -13.2$\pm$11.8 & -3.6$\pm$6.2 & -12.9$\pm$6.3 & 140.1$\pm$17.8\\ 
94494 & 0.7401 & 10.9 & 0.49$\pm$0.00 & 0.95$\pm$0.00 & 39.50$\pm$0.30 & 3.72$\pm$0.05 & 92.0$\pm$1.0 & 63.1$\pm$0.7 & 165.3$\pm$1.4 & 215.5$\pm$3.9\\ 
97994 & 0.9821 & 11.2 & 0.57$\pm$0.01 & 0.64$\pm$0.01 & 19.79$\pm$0.48 & 2.61$\pm$0.06 & 115.8$\pm$6.1 & 83.6$\pm$4.4 & 189.3$\pm$8.9 & 236.6$\pm$17.6\\ 
105208 & 0.9345 & 10.8 & 0.64$\pm$0.02 & 0.85$\pm$0.01 & 31.85$\pm$0.68 & 5.74$\pm$0.20 & -31.1$\pm$10.5 & -27.4$\pm$9.3 & -55.3$\pm$20.7 & 203.0$\pm$22.9\\ 
107468 & 0.9178 & 11.1 & 0.21$\pm$0.00 & 0.24$\pm$0.00 & 24.21$\pm$0.35 & 2.27$\pm$0.03 & -97.8$\pm$3.0 & -15.1$\pm$1.1 & -122.4$\pm$1.9 & 234.0$\pm$6.2\\ 
107489 & 0.8383 & 11.1 & 0.32$\pm$0.00 & 0.44$\pm$0.00 & 26.75$\pm$0.26 & 2.29$\pm$0.03 & 29.8$\pm$4.0 & 7.6$\pm$1.0 & 60.5$\pm$2.0 & 383.6$\pm$7.0\\ 
108227 & 0.9603 & 11.4 & 1.66$\pm$0.03 & 0.55$\pm$0.01 & -31.66$\pm$0.54 & 1.30$\pm$0.03 & -28.0$\pm$5.7 & -43.5$\pm$9.8 & -59.2$\pm$24.0 & 263.9$\pm$17.0\\ 
108472 & 0.6671 & 10.6 & 0.14$\pm$0.00 & 0.56$\pm$0.01 & 8.52$\pm$0.68 & 3.72$\pm$0.07 & -69.5$\pm$2.1 & -12.2$\pm$0.4 & -139.8$\pm$0.8 & 160.8$\pm$4.2\\ 
110509 & 0.6671 & 11.0 & 0.99$\pm$0.01 & 0.95$\pm$0.00 & 33.81$\pm$0.23 & 3.76$\pm$0.04 & 23.9$\pm$1.5 & 29.6$\pm$2.0 & 41.6$\pm$5.8 & 217.4$\pm$5.0\\ 
110805 & 0.7292 & 10.6 & 0.47$\pm$0.00 & 0.21$\pm$0.00 & 12.40$\pm$0.22 & 0.55$\pm$0.01 & -151.5$\pm$3.0 & -70.4$\pm$2.7 & -178.8$\pm$2.3 & 172.7$\pm$7.5\\ 
111188 & 0.9164 & 10.9 & 0.43$\pm$0.01 & 0.58$\pm$0.01 & 11.23$\pm$0.65 & 5.62$\pm$0.19 & -53.3$\pm$9.5 & -46.3$\pm$5.7 & -55.7$\pm$13.5 & 180.1$\pm$10.1\\ 
112200 & 0.8279 & 10.6 & 0.29$\pm$0.01 & 0.86$\pm$0.01 & 36.97$\pm$0.77 & 4.21$\pm$0.15 & 3.5$\pm$8.5 & 1.3$\pm$5.1 & 3.2$\pm$7.1 & 151.6$\pm$13.7\\ 
112534 & 0.9837 & 11.0 & 0.34$\pm$0.01 & 0.50$\pm$0.01 & -21.17$\pm$0.60 & 1.89$\pm$0.06 & -107.5$\pm$7.8 & -45.0$\pm$5.5 & -159.7$\pm$7.3 & 297.3$\pm$19.9\\ 
116829 & 0.6683 & 10.8 & 0.45$\pm$0.00 & 0.70$\pm$0.00 & 22.15$\pm$0.29 & 2.46$\pm$0.03 & -1.5$\pm$2.0 & -1.5$\pm$1.7 & -1.5$\pm$2.0 & 162.3$\pm$6.7\\ 
117010 & 0.6766 & 10.4 & 0.40$\pm$0.01 & 0.54$\pm$0.01 & -27.19$\pm$0.61 & 4.45$\pm$0.13 & 99.0$\pm$3.4 & 36.3$\pm$1.8 & 135.3$\pm$2.7 & 158.1$\pm$5.3\\ 
117400 & 0.6687 & 11.3 & 0.80$\pm$0.01 & 0.79$\pm$0.00 & 40.78$\pm$0.28 & 4.85$\pm$0.06 & 1.3$\pm$3.7 & 1.2$\pm$3.5 & 3.1$\pm$5.3 & 258.2$\pm$4.7\\ 
117692 & 0.6753 & 10.8 & 0.63$\pm$0.01 & 0.48$\pm$0.00 & -22.70$\pm$0.34 & 4.13$\pm$0.07 & -100.9$\pm$2.4 & -54.2$\pm$1.4 & -189.4$\pm$2.6 & 185.4$\pm$7.3\\ 
\enddata
\tablecomments{Table 1 will be published in its entirety in the machine-readable format. A portion is shown here for guidance regarding its form and content.  This table includes measured properties of the galaxies included in this sample from the year one LEGA-C dataset. All galaxies included in this table are well-aligned with the N-S VIMOS slits ($\mathrm{|PA|<45^o}$), quiescent based on \citet{muzzin:13} U-V and V-J rest-frame color cuts, have reliable morphological parameters measured from ACS F814 images, and represent single Virialized systems. Columns: (1) ID from the \citet{muzzin:13ultravista} UltraVISTA DR1 v4.1 catalogs; (2) spectroscopic redshift; (3) log Stellar Mass assuming \citet{chabrier:03} IMF; (4) S\'ersic semi-major axis; (5) projected axis ratio; (6) major axis position angle; (7) S\'ersic index; (8) average line-of sight rotational velocity measured at 5kpc; (9)  average line-of sight rotational velocity measured at the effective radius; (10) average line-of sight rotational velocity measured at the maximum extent; (11) velocity dispersion measured in the central pixel in the spatial dimension.}
\end{deluxetable*}

These fits yield spatially resolved line-of-sight stellar and gas velocity and velocity dispersion profiles along the N-S slits. Although emission lines are present due to residual ionized gas (primarily the $\mathrm{[OII]}$ doublet) in a subset of this quiescent galaxy sample, we focus our analysis on the kinematics derived from fitting the stellar continuum of each galaxy. Measured stellar rotation curves are shown in Figure \ref{fig:rot}, in which velocity of the stellar component is indicated by black points and stellar velocity dispersion profiles are shown in red.  Rotation curves are plotted in order of increasing rotational velocity, separated by page in decreasing mass bins.  We fit the rotation curves with an arctangent model and define the line-of-sight rotational velocity ($V_5$) of a galaxy as the value of the best-fitting arctangent at a radius of 5 kpc { along the slit. This distance is not corrected for inclination or slit misalignment.} We define the central velocity dispersion ($\sigma_0$) as the velocity dispersion measured in the central pixel ($0.205"$){, which is set as the brightest pixel in the spatial profile}. Uncertainties in $V_5$ are estimated by bootstrap resampling within the velocity errors and errors in velocity dispersion are formal uncertainties estimated by PPXF, with a small correction to underestimated formal errors based on the measured relationship between the measured S/N and formal errors. 

We adopt this definition of rotational velocity within a fixed physical aperture for two primary reasons. First, the effects of seeing will be similar within a fixed physical radius as opposed to an aperture that scales with the galaxy size. The 5 kpc aperture is used because it is the approximate extent of the shortest LEGA-C rotation curves, and therefore requires minimal extrapolation. Secondly, utilizing a fixed aperture allows for comparison with galaxies at low redshift in \S \ref{sect:califa} within the same physical region of the galaxy and will be less sensitive to differing apertures due to real size evolution in the galaxy populations. We discuss the impact of this choice of aperture, including the effects of adopting an evolving aperture {or utilizing the maximum observed velocity}, in the Appendix \ref{sect:V}.

Given that the effective seeing, including atmospheric and alignment effects, is comparable to the spatial extent of the galaxies themselves (FWHM ${\sim}1.0''{\approx}7$ kpc) the effects of beam smearing will be significant and kinematic measurements at each pixel ($0.205"$) are not independent. This results in shallower than intrinsic rotation curves and elevated line-of-sight velocity dispersions.
Dynamical modeling that accounts for aperture and beam smearing effects, which is common in the analysis of emission line kinematics at high redshift \citep[e.g.,][]{vogt:96,vogt:97,weiner:06,kassin:07,simons:15,simons:16,price:16,simons:16,wuyts:16,harrison:17,straatman:17} can reconstruct the intrinsic rotation and velocity dispersion profiles, given modeling assumptions, for direct comparisons with present-day galaxy samples. Such modeling efforts are in preparation { (van Houdt et al.\ in prep)}, but beyond the scope of the current paper; here we focus on the directly measured rotation ($V_5$) and rotational support ($|V_5| / \sigma_0$). In \S \ref{sect:califa} we reconstruct the rotation and dispersion profiles of local galaxies as they would be observed with LEGA-C at $z\sim 0.8$.

\section{Stellar Rotation in Quiescent LEGA-C Galaxies}\label{sect:results}

\begin{figure*}[!t]
    \centering
    \includegraphics[width=0.85\textwidth]{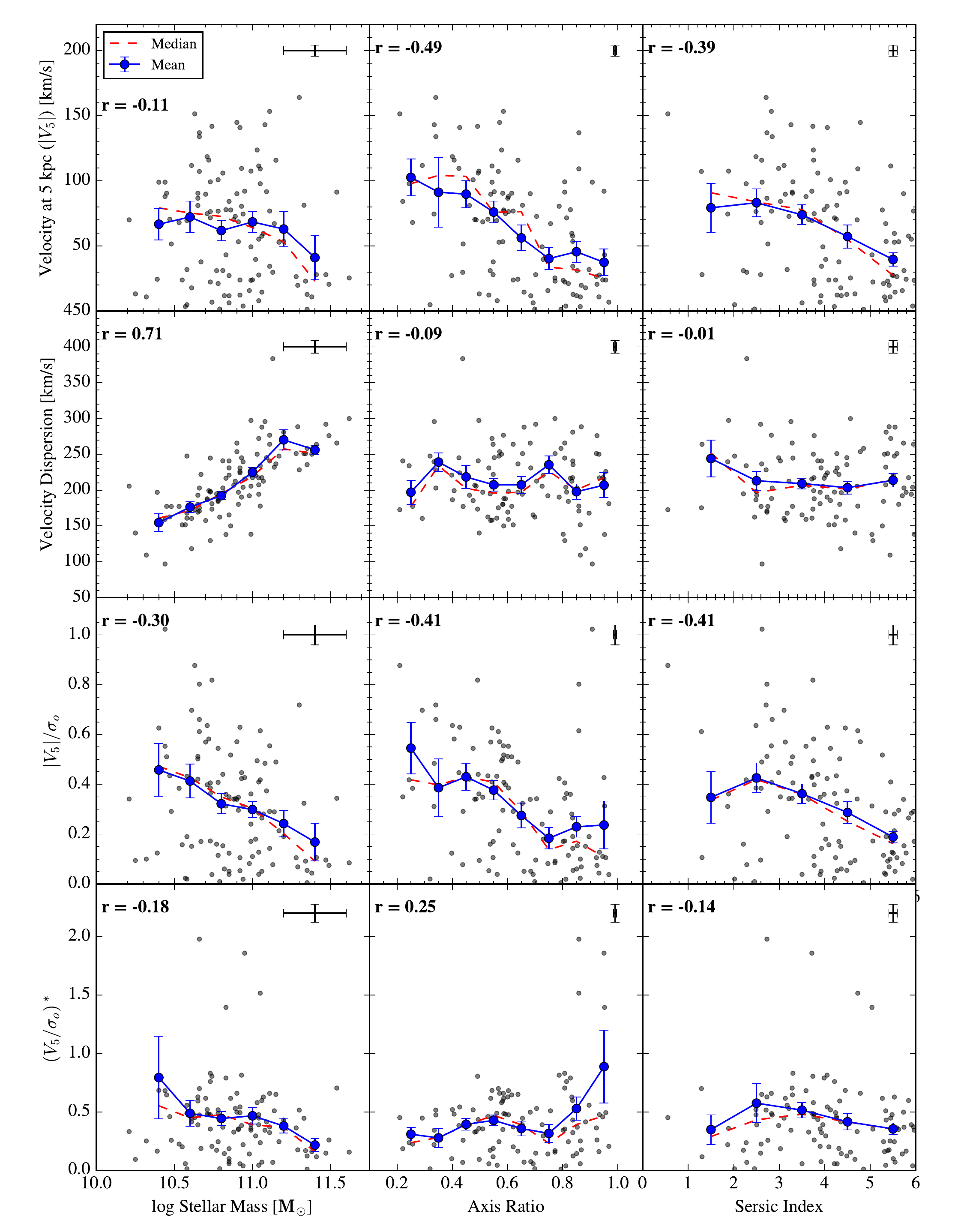}
    \caption{Rotational velocity (${|V_5|}$, top row), { central velocity dispersion ($\sigma_0$, second row),} rotational support ($|V_5|/\sigma_0$, { third} row), and rotational support normalized by the expectation for an oblate rotator given the measured projected axis ratio ($(V_5/\sigma_0)*$, bottom row) in quiescent LEGA-C galaxies versus stellar mass (left), projected axis ratio (middle), and S\'ersic index (right). Individual galaxies are indicated by small gray symbols, median and mean trends are indicated by red dashed and blue solid lines and symbols, respectively.  { The strongest correlation exists between stellar mass and velocity dispersion, or the ``mass'' Faber-Jackson relation.} Projected axis ratio exhibits the strongest anti-correlation with $|V_5|/\sigma_0$ and unlike S\'ersic index, the population average with $|V_5|/\sigma_0$ does not flatten out at elongated axis ratios in this sample. When projection effects are minimized with $(V_5/\sigma_0)^*$, this removes significant correlations with projected axis ratios, suggesting roughly similar correlations between rotational support and stellar mass, axis ratio, and S\'ersic index.}
    \label{fig:vsig_trends}
\end{figure*}

In this section we investigate trends of stellar rotation and rotational support with other properties of massive quiescent galaxies. We specifically focus on stellar mass, with which rotational support has been demonstrated to depend in $z\sim0$ galaxies, and on the two photometric measures that have been used to assess the ``disk-like'' nature of massive quiescent galaxies at high redshifts: projected axis ratio and S\'ersic index \citep[e.g.,][]{wel:11,chevance:12,chang:13,cappellari:16,graham:18}. { We note that the measured $|V|/\sigma$ will depend on projection effects, which is particularly important in interpreting trends in projected axis ratio. Therefore in addition to $|V_5|/\sigma_0$, we introduce $(V_5/\sigma_0)^*$ following e.g. \citet[][]{binney:78,davies:83}, which is defined as the $V/\sigma$ normalized by the $(V/\sigma)_O$ for an oblate, isotropic model and should be largely independent of projection effects. We adopt the approximation $V/\sigma\approx \sqrt{\epsilon/(1-\epsilon)}$ from \citet{kormendy:82a}. {Following this definition,}

\begin{equation}
(V_5/\sigma_0)^* = \frac{(|V_5|/\sigma_0)}{\sqrt{\epsilon/(1-\epsilon)}}.
\end{equation}

Figure \ref{fig:vsig_trends} shows { rotational velocity ($V_5$)} of galaxies in the top row, velocity dispersion in the second row, rotational support ($|V_5|/\sigma_0$) in the third row, and in the bottom row rotational support with a correction for projection effects, $(V_5/\sigma_0)^*$, as a function of stellar mass (left), projected axis ratio (center), and S\'ersic index (right)}. Average uncertainties on the measurements are indicated by errorbars in the upper right corners of each panel. Running median and mean are indicated by red dashed and blue solid lines respectively for bins with greater than three data points. Errors on the mean are estimated in each bin via jackknife resampling. In each case these trends are best described as scatter between no rotational support and a maximum value that depends on the property plotted on the horizontal axis. This leads to measured (anti-)correlations, for which we quote the Pearson correlation coefficient in the upper left corner of each panel.

\begin{figure}[t]
    \centering
    \includegraphics[width=0.44\textwidth]{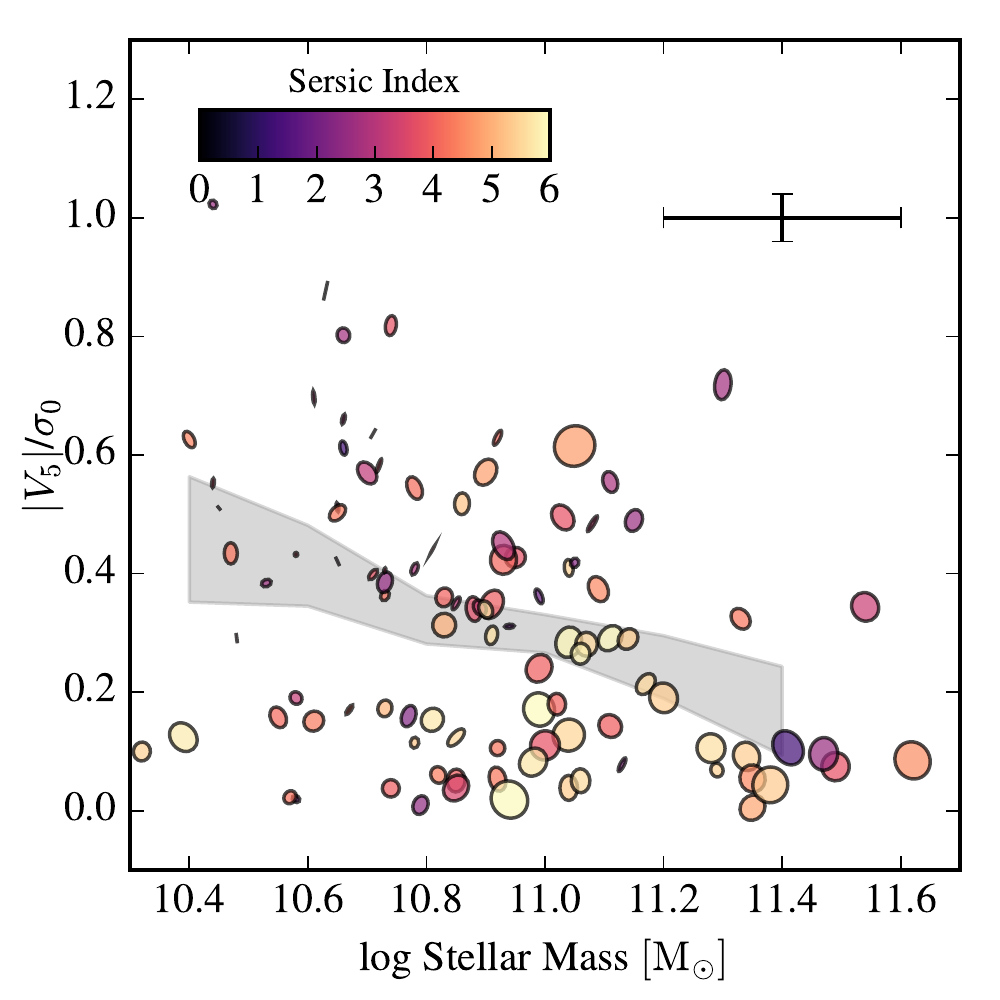}
    \caption{Rotational support ($|V_5|/\sigma_0$) versus stellar mass for the LEGA-C sample of massive, quiescent galaxies. Symbol size indicates the galaxy effective radii (in log scale) and position angles and axis ratios of symbol ellipses reflect those of the galaxies. Symbol color indicates S\'ersic index. The mean relation is indicated by gray band and average uncertainty is indicated by errorbars in the upper right corner. The majority of high-mass galaxies have minimal rotational support, even when their S\'ersic indices are disk-like, however there are several high-mass galaxies with significant rotation. Below $\log M/M_{\odot}\lesssim11.2$ galaxies exhibit a range in rotational support and smaller and more elongated galaxies consistently show higher measured $|V_5|/\sigma_0$.}
    \label{fig:vsig_lmass_qn}
\end{figure}

In the left panels, we see that more massive galaxies { exhibit lower rotational support ($|V_5|/\sigma_0$ or $(V_5/\sigma_0)^*$) than less massive galaxies}. This is also evident in the local universe \cite[e.g.,][]{emsellem:11}. We will return to this trend in Figure \ref{fig:vsig_lmass_qn}, where we also include information about galaxy morphology in the same panel. We emphasize that this is primarily due to the known correlation between stellar mass and velocity dispersion, the mass \citet{faberjackson} relation { (left panel, second row)}; rotational velocities alone do not exhibit a strong correlation with stellar mass (top left panel). However, at all masses there is at least a small fraction of galaxies that are observed to have very little rotational support. Some of this is an observational effect: beam smearing, { inclination}, and slit misalignment diminish ordered rotation and increase observed velocity dispersions and we expect this to preferentially impact smaller galaxies. We investigate these effects in greater detail in Section \ref{sect:califa_sims}.

Another key result of our measurements is that galaxies that are flat in projection generally show rotation in their stellar body, whereas round galaxies do not (top center panel in Figure \ref{fig:vsig_trends}). This is well-understood as largely due to a combination of intrinsic elongation and projection effects \citep[e.g.,][]{cappellari:07,emsellem:07,emsellem:11,fogarty:14,fogarty:15,sande:17}. This trend is tightened when rotational support is compared to dispersion support in the central pixel ({ third row, center panel} in Figure \ref{fig:vsig_trends}), with a Pearson correlation coefficient $r=-0.41$. { This is primarily a trend in rotational velocity, not velocity dispersion (see middle panel, second row)}. There is a subset of elongated galaxies that show little rotation ({ 3} of 25 galaxies with $b/a<0.5$ have $|V_5|/\sigma_0<0.1$). The nature of these galaxies remains to be determined, but perhaps they are not unlike NGC4550, which does not show net rotation but has been demonstrated to consist of two counter-rotating disks \citep{johnston:13}. This overall trend implies that the distribution of projected axis ratios for a population of { quiescent} galaxies will be a decent estimate of the overall { observed} degree of rotational support. However, for any individual galaxy with an observed axis ratio of $b/a\gtrsim0.6$ a significant fraction of galaxies will still have significant rotation and spatially resolved kinematics will be necessary to distinguish between pressure and rotationally supported systems.

Intriguingly, although both velocity (top right { panel}) and rotational support ({ third row, right panel}) exhibit a statistically significant correlation with S\'ersic index, the mean relation turns over exactly at the S\'ersic index where one would expect the anti-correlation to be strongest.  Although the numbers are small, the mean rotational velocity of galaxies that would be classified as disk-like based on their concentrations ($n<2.
5$) is not elevated ($\langle|V_5|/\sigma_0\rangle=0.33$, median={ 0.34}) compared to the overall average ($\langle|V_5|/\sigma_0\rangle=0.31$). This trend is strongest for the highest mass quiescent galaxies ($\log M_{\star}/M_{\odot} >11$), for which the $n<2.5$ average $\langle|V_5|/\sigma_0\rangle=0.20$ versus overall $\langle|V_5|/\sigma_0\rangle=0.25$. These massive galaxies are the most extended, and therefore the least affected by beam smearing, and yet this trend is contrary to expectations. Larger samples, such as the full 4 year LEGA-C sample, will likely include a larger number of $n<2.5$ galaxies and allow for a more statistically significant assessment of these trends. Regardless, we emphasize that measuring the S\'ersic index of an individual quiescent galaxy cannot determine whether it is rotationally supported. Overall, S\'ersic index is anti-correlated with rotational support, with a weaker Pearson coefficient $r=-0.41$).

{ Although measured $|V|/\sigma$ will likely be sensitive to projection effects, $(V_5/\sigma_0)^*$, which normalizes out expected $V/\sigma$ based on projected axis ratios for an oblate, isotropic model, should be largely independent of projection effects. The bottom row of Figure \ref{fig:vsig_trends} shows} $(V_5/\sigma_0)^*$ as a function of stellar mass, projected axis ratio, and S\'ersic index in the bottom panels. Although all quantities are still correlated with this measure of rotational support, it is clear that a significant fraction of the correlation with projected axis ratio was covariance of the variables; once the projection effects are removed { projected axis ratio exhibits a mild correlation with rotational support ($r=0.25$). This remaining correlation is likely driven by the four round $(b/a > 0.8)$ galaxies with high $(V_5/\sigma_0)^*$ that are not well approximated by isotropic oblate rotators}. { We note that although inclination and projection effects can account for some of the anti-correlation between $|V_5|/\sigma_0$ and S\'ersic index, the weak anti-correlation remains between $(V_5/\sigma_0)^*$ and S\'ersic n. We reiterate that this sample includes very few low S\'ersic index galaxies and although we caution again the use of S\'ersic index to characterize individual galaxies, we do not have the statistics to characterize this trend at low S\'ersic indices.}

Figure \ref{fig:vsig_lmass_qn} shows rotational support ($|V_5|/\sigma_0$) versus stellar mass, but now with symbols that reflect morphologies. Symbol sizes correspond to logarithmically scaled galaxy effective radii, symbol axis ratios and position angles reflect the projected galaxy shapes and orientations. Symbol colors correspond to S\'ersic index. The average uncertainty is indicated by the errorbars in the upper right and the mean trend, as calculated in Figure \ref{fig:vsig_trends}, is indicated by the gray band. Here we can clearly identify massive galaxies with seemingly inconsistent morphologies and measured kinematics: galaxies with little observed rotational support, but low S\'ersic indices (purple colors) as well as others with high S\'ersic indices (orange colors) and high $|V_5|/\sigma_0$. 

Our measured $|V_5|/\sigma_0$ is likely to be an underestimate due to a number of observational effects such as rotational velocities contributing to central velocity dispersions and decreasing measured line-of-sight velocities due to inclination. Therefore, galaxies with low measured rotational support may in fact be revealed to be intrinsically fast-rotators with full modeling; however galaxies that are observed to be rotating quickly cannot be slow-rotators. Given this observational ambiguity we refrain from using the terms ``fast'' and ``slow'' rotators, but return to quantifying the observational biases in the following section.

Our kinematic measure $|V_5|/\sigma_0$ is not directly comparable to the classifiers used for present-day galaxies as seeing, slit-misalignment, and other observational effects are not taken into account. However, the trends in Figure \ref{fig:vsig_trends} are very similar to those observed for present-day galaxies \citep[e.g.,][]{emsellem:11}, and we conclude that at all cosmic times since at least $z\sim 1$ the quiescent galaxy population consists of galaxies with low and high degrees of rotational support that reflect their intrinsic structure (spheroidal/triaxial and disk-like/oblate, respectively). 
At the same time, among the 10 most massive galaxies with stellar masses $>2\times10^{11}M_{\odot}$, only 2 show evidence for rotation. This is suggestive that the only way that galaxies can grow to such large masses is by a mechanism that reduces the angular momentum, that is, dissipationless merging. In the following section, we analyze the CALIFA dataset to further explore the question of quantifying this evolution.

\begin{figure}[!t]
    \centering
    \includegraphics[width=0.45\textwidth]{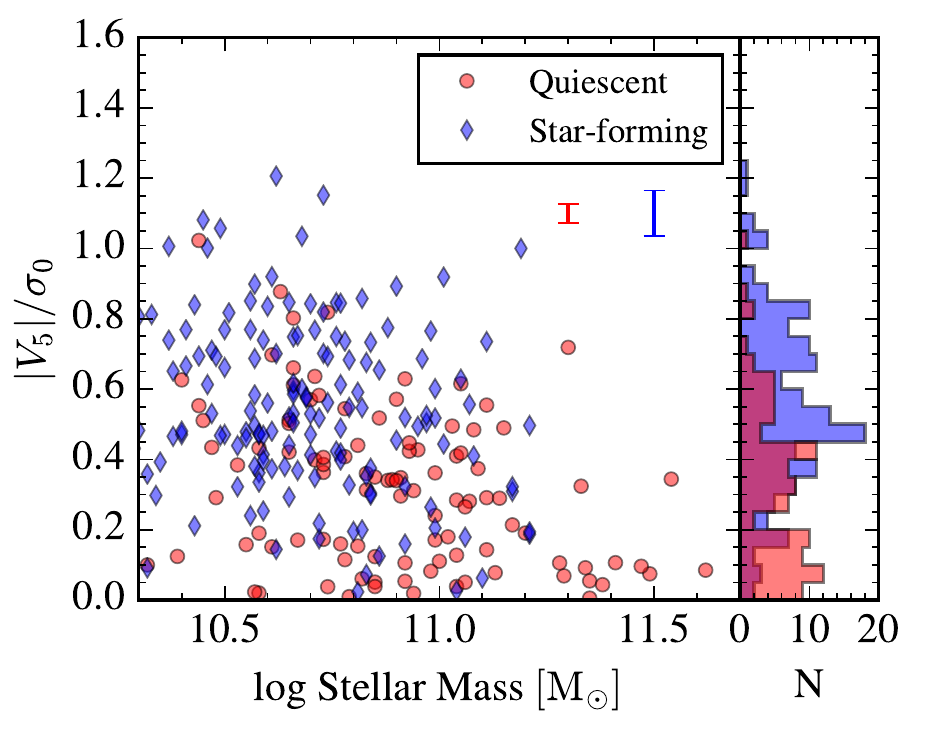}
    \caption{Observed rotational support of LEGA-C galaxies versus stellar mass for star-forming (blue diamonds) and quiescent galaxies (red circles). Average uncertainties, shown as blue and red errorbars in the upper right, are higher for the star-forming galaxies ($\sim0.1$) than for quiescent galaxies ($\sim0.04$) in the LEGA-C sample. The right panel indicates the histograms in rotational support between the star-forming and quiescent populations; the distributions are overlapping but on average star-forming galaxies show higher $V_5/\sigma_0$ than quiescent galaxies overall and at fixed mass.}
    \label{fig:vsig_sfq}
\end{figure}

Although we focus on the quiescent sample only for this paper, we note that as expected, the star-forming and quiescent galaxy populations differ in dynamics as well as stellar populations. Figure \ref{fig:vsig_sfq} shows the observed rotational support ($|V_5|/\sigma_0$) versus stellar mass for all galaxies with photometric axes within 45$^o$ of the N-S slits. Quiescent galaxies are indicated by red circles and star-forming galaxies by blue diamonds. The star-forming galaxies have more rotational velocity than quiescent galaxies, as found in the local Universe \citep[e.g.][]{cortese:16}. Figure \ref{fig:sample} demonstrates the known bimodality of these two populations in size and specific star formation rate, this figure provides the first evidence for dynamical bimodality at high redshift based on stellar kinematics.  The two populations overlap in observed phase space, however their distributions differ significantly (see histograms in the right panel).  A two-sample K-S test rejects the possibility that they are drawn from the same distribution with a ${p=1\times10^{-10}}$, or ${p=4\times10^{-8}}$ for massive $\log M_{\star}/M_{\odot} > 10.4$ galaxies. Average values of errors on $V_5/\sigma_0$ for the star-forming and quiescent sub-samples are indicated by blue and red errorbars in the upper right corner. Uncertainties in the $|V_5|/\sigma_0$ values, especially for the star-forming population, contribute significantly to the broadening of the distribution. Therefore, this discrepancy may be stronger in the intrinsic properties of the two populations. We leave the analysis of the dynamics of star-forming galaxies and of the joint population to future studies (Straatman et al. in prep, van Houdt et al. in prep).

\section{CALIFA stellar kinematics and the redshift evolution of rotational support}\label{sect:califa}

The Calar Alto Legacy Integral Field Area (CALIFA) survey provides an excellent census of the spectroscopic properties of local ($0.005<z<0.03$) galaxies of all morphological and spectral types \citep{sanchez:12,walcher:14}. The CALIFA team has promptly provided reduced data products in public data releases in addition to derived spectroscopic properties. For this project we include CALIFA galaxies from Data Release 3 \citep[DR3,][]{sanchez:16}, stellar kinematics maps from \citet{falconbarroso:17}, and spectroscopic classifications based on ionized gas lines from \citet{canodiaz:16}. Using this dataset, we use intensity, stellar velocity, and stellar velocity dispersion fields in two spatial dimensions and extract profiles along a variety of axes and replicate the LEGA-C kinematic analysis on a local sample, quantifying intrinsic properties and simulating the effects of seeing on the measured LEGA-C rotation curves. 
\begin{figure}[t]
    \centering
    \includegraphics[width=0.45\textwidth]{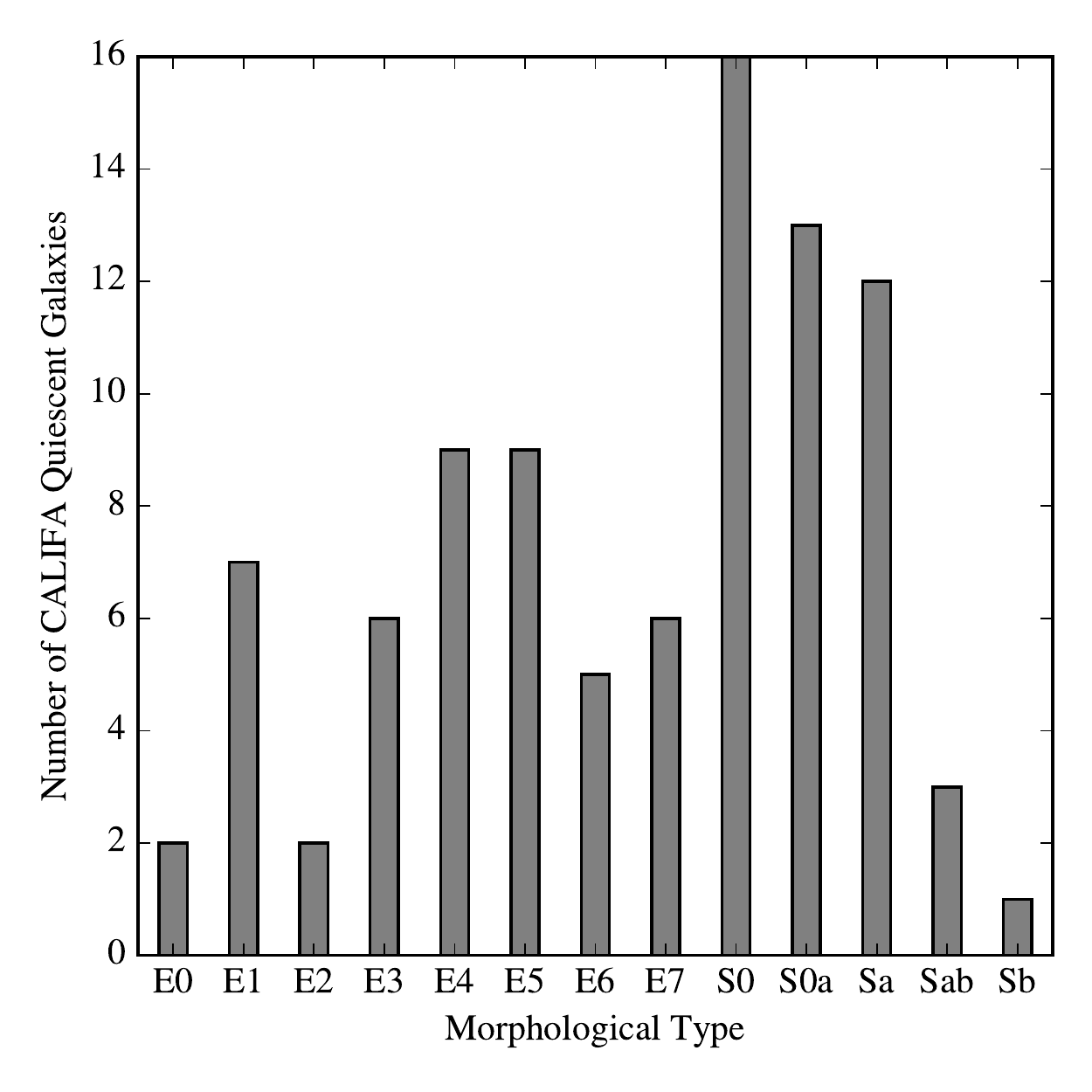}
    \caption{Morphological distribution of the 91 CALIFA galaxies determined to be quiescent based on $\mathrm{EW(H\alpha)<3\AA}$ in the stellar kinematics sample.}
    \label{fig:califa_morph}
\end{figure}

\subsection{The CALIFA Dataset}

Of the 667 galaxies in the full DR3, 300 are included in the \citet{falconbarroso:17} analysis of stellar kinematics. This sample of galaxies, which have been observed with both low (V500) and medium (V1200) resolution gratings, is deemed to be representative of the full CALIFA sample in magnitude, size, and redshift and spans a wide range of morphological types. As in the LEGA-C sample, \citet{falconbarroso:17} remove strongly interacting galaxies from this kinematic sample.  \citet{falconbarroso:17} analyze IFU datacubes for each galaxy, which are Voronoi binned to $S/N\sim20$ and the stellar kinematics are measured in each bin by fitting a combination of stellar templates convolved with a gaussian line-of-sight velocity dispersion. These fits yield maps of velocity and velocity dispersion at each spaxel, which the authors provide on the CALIFA website (\url{http://califa.caha.es/?q=content/science-dataproducts}). \citet{falconbarroso:17} also provide stellar masses assuming a \citet{chabrier:03} IMF and effective radius, ellipiticity, and position angle determined from the outer parts of the galaxies in SDSS imaging as described in \citet{walcher:14}. 

We further limit our analysis to quiescent galaxies following the classifications of \citet{canodiaz:16}, who determine $H\alpha$-based star formation rates and use ionized gas lines to differentiate amongst dominant ionization sources using $\mathrm{EW(H\alpha)}$ and the Kewley demarcation limit \citep{kewley:01} in the Baldwin-Philips-Terlevich (BPT) diagram \citep{baldwin:81}. \citet{canodiaz:16} identify each CALIFA galaxy as either ``Star-forming'', ``AGN'',``Retired'', or in ambiguous cases ``Undefined''. The \citet{canodiaz:16} study included a representative sample of 535 galaxies that had been observed by February, 2015; and therefore does not completely overlap with the \citet{falconbarroso:17} sample. We classify the remaining ten galaxies by eye using the two-dimensional star-formation maps provided in the CALIFA DR3. For the most conservative comparison with the current study, we limit our analysis to the quiescent or ``Retired'' ($\mathrm{EW(H\alpha)<3\AA}$) sample of galaxies, based on their spectroscopic properties. { Only 4 galaxies are classified as Retired by eye and we verify that excluding these galaxies does not significantly impact any of the conclusions in this paper.} For maximum consistency in stellar population modeling, we compare stellar masses with those derived by \citet{brinchmann:04} for the subset of these galaxies which also fall in the spectroscopic SDSS DR7 sample. These fits are also based on aperture photometry and are analyzed using similar methodology to our modeling of the UltraVISTA photometry. We find that CALIFA stellar masses are higher than those derived by \citet{brinchmann:04} by a median of 0.16 dex for the retired galaxy population. We perform a linear regression to this subset and apply this correction to the CALIFA-derived stellar masses. The final sample includes 91 galaxies across a range of morphological types, from E0 to Sb as shown in Figure \ref{fig:califa_morph}.

\begin{figure*}[t]
    \centering
    \includegraphics[width=\textwidth]{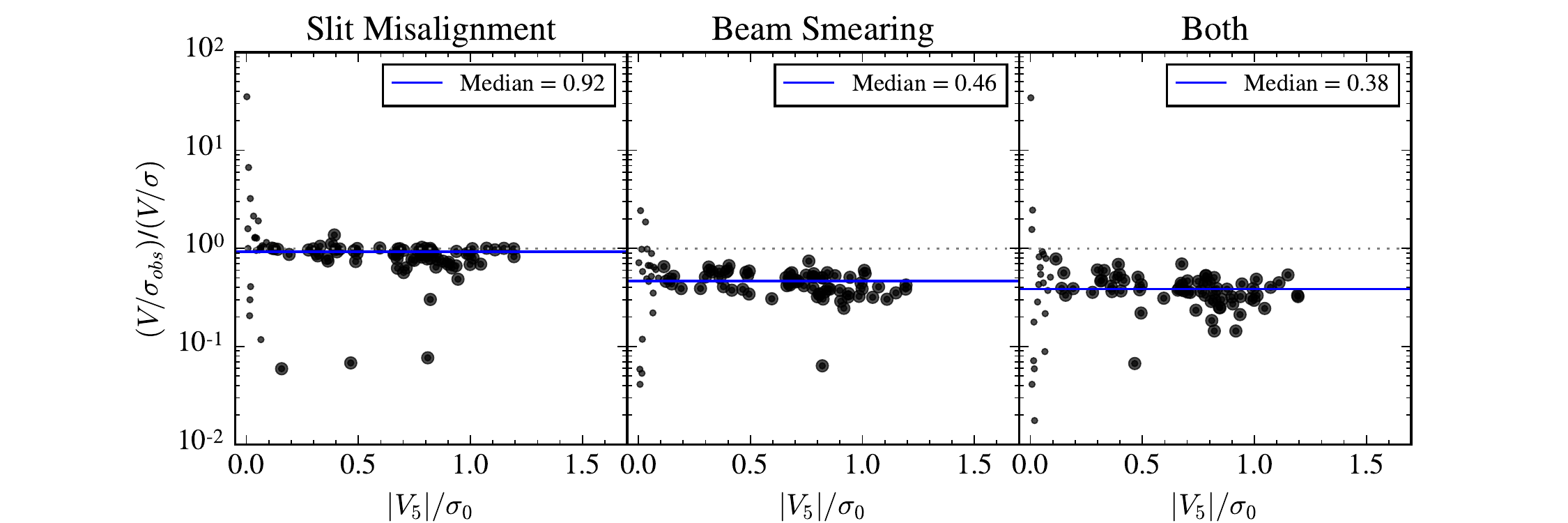}
    \caption{Ratio of simulated ``observed'' { to intrinsic rotational support ($|V_5|/\sigma_0$) versus the intrinsic value} due to slit misalignment (left panel), beam smearing (center panel), and the combined effects (right panel). { Galaxies with minimal rotational support ($|V_5|/\sigma_0<0.1$) are indicated by small symbols and those with higher $|V_5|/\sigma_0$ by large symbols}. Beam smearing is the dominant effect, decreasing the observed $|V_5|/\sigma_0$ by an { median} factor of ${\sim}2.2$, while slit misalignment decreases the value by ${\sim}8\%$.}
    \label{fig:obs_effects}
\end{figure*}

\begin{figure*}[]
    \centering
    \includegraphics[width=0.8\textwidth]{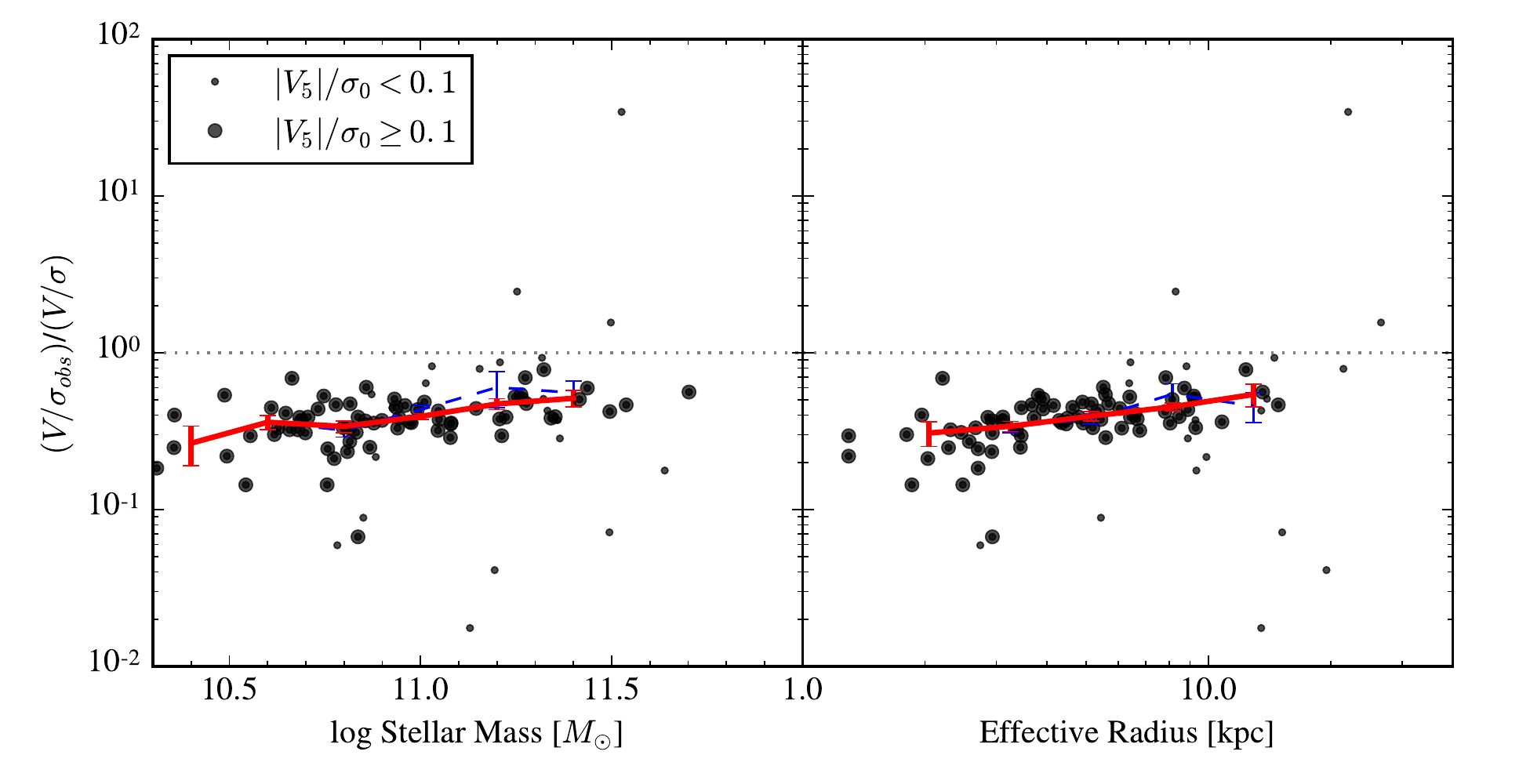}
    \caption{Trends in the ratio of observed to intrinsic $|V_5|/\sigma_0$ from the CALIFA simulations with stellar mass in the left panel and effective radius in the right panel. Small symbols indicate galaxies with $|V_5|/\sigma_0<0.1$, which are most sensitive to this relative metric. The running mean relations are indicated by blue dashed (all galaxies) and red solid ($|V_5|/\sigma_0>0.1$) lines. As expected, the blurred rotational support preferentially impacts the least massive and most compact galaxies because of the relative size of the PSF and the galaxy extent.}
    \label{fig:califa_trends}
\end{figure*}

\subsection{Simulating LEGA-C observations with CALIFA datacubes}\label{sect:califa_sims}

For each galaxy in the quiescent CALIFA sample, we extract the intrinsic intensity $I_{1D}(x)$, velocity $V(x)$, and velocity dispersion $\sigma(x)$ profiles along lines passing through the maximum of the intensity map of the galaxy. These 1D profiles are measured along the published galaxy photometric position angles, as determined by \citet{walcher:14} from galaxy outskirts in the SDSS imaging. Rotation curves are fit with arctangent functions and rotational velocity at 5 kpc and central velocity dispersion are measured as for the LEGA-C dataset. These values correspond to the intrinsic $V_5$ and $\sigma_0$ values.

We use the two-dimensional intensity and kinematic maps, spatially subsampled by a factor of 100, to simulate the observational effects of the misaligned 1'' slits ($\sim7.5$kpc), 0.205'' pixels ($\sim1.5$kpc), and seeing characteristic of the LEGA-C observations. Slit misalignment in the LEGA-C survey, which in this study is limited to within 45$^o$ of the North-South slits, is simulated by extracting one-dimensional profiles along the closer of the horizontal or vertical directions. The intensity in two-dimensional position-velocity space can be defined as:

\begin{equation}
I_{3D}(x,y,v) = I_{2D}(x,y)\exp\Bigg[-\frac{(v-V(x,y))^2}{2\sigma(x,y)^2}\Bigg].
\end{equation}

The effects of seeing are then simulated by convolving this intensity ($I_{3D}$) field with a two-dimensional Moffat profile. For this we adopt a uniform value of $\mathrm{FWHM_{PSF}}=7\mathrm{kpc}$ as representative of the LEGA-C redshift and spectrum ($\mathrm{FWHM}\sim1.0''$ at the average z=0.78). We adopt a value of $\beta=4.765$ following \citet{trujillo:01}. The velocity and velocity dispersion profiles are then measured as the intensity-weighted first and second moments of $I_{3D}(x,y,v)$, summed within a 7.5kpc band perpendicular to the horizontal or vertical slit and within 1.5kpc pixels along the slit. These profiles are measured from the initial and convolved intensity fields. The ``observed'' rotational velocity ($V_5$) and central velocity dispersion ($\sigma_0$) are measured as in the LEGA-C dataset.  The resulting intrinsic and binned unconvolved and convolved rotation curves and velocity dispersion profiles of the CALIFA galaxies are included in Figure \ref{fig:rot_califa} in Appendix \ref{sect:califa_rot}.

Figure \ref{fig:obs_effects} shows the ratio of simulated ``observed'' $|V_5|/\sigma_0$ to the intrinsic value for the two main effects included in the simulation, slit misalignment (left panel) and beam smearing (center panel), and for the combination versus the intrinsic value measured along the position angle (right panel). In each panel the median value is indicated by a horizontal blue line. Overall, the impact of slit misalignment and beam smearing from the simulated PSF ($\mathrm{FWHM_{PSF}}=7\mathrm{kpc}$) on the measured rotation curves is significant, with the latter dominating the difference from the intrinsic and blurred 1D rotation curves. Slit misalignment decreases the measured ratio by { an} average of ${\sim}8\%$.  { \citet{straatman:17} found the impact of slit misalignment to be stronger for emission line galaxies, finding that this effect decreases the measured velocities by a median factor of 1.19 with significant scatter. However, we note that those simulations were for a very different sample of galaxies and were produced using infinitely thin galaxy models. It may be the case that at $z\sim1$ the LEGA-C quiescent galaxies are more disk-like than quiescent galaxies in CALIFA \citep[e.g.][]{chang:13}, however these galaxies will likely be either triaxial or oblate spheroids and not well described by thin disk models. These thin disk models would over-predict the effect of slit misalignment for a sample with likely non-zero minor axis rotation.}

The second effect of beam smearing (center panel) is driven by differences in the measured rotational velocity as ordered motion contributes to velocity dispersion in the outer parts of the simulated galaxies. The measured $|V_5|/\sigma_0$ decreases by an average factor of ${\sim}2.5$ after convolution with a $7$kpc PSF, and this effect would only increase with a larger PSF. Although beam smearing significantly changes the velocity dispersion profiles, it only minimally influences the measured central velocity dispersion, with a median ratio of observed velocity dispersion to intrinsic of 0.98 and in all cases it is less than a ${\sim}\,10\%$ effect. Therefore the diminished $|V_5|/\sigma_0$ in Figure \ref{fig:obs_effects} is primarily due to the lowered $|V_5|$.

Beam smearing impacts smaller galaxies more severely than large galaxies, as shown in Figure \ref{fig:califa_trends}. This figure shows the ratio of simulated to intrinsic $|V_5|/\sigma_0$ versus galaxy stellar mass and size in the CALIFA sample. In each panel the running average is indicated by the blue dashed line for the full sample and red solid line for galaxies with $|V_5|/\sigma_0>0.1$, for which uncertainties are measured via jackknife resampling. Measured rotational support will be reduced by nearly an order of magnitude for the lowest mass and smallest galaxies in the sample, whereas the largest and most massive galaxies are less impacted by these simulations. These trends suggest that the differential effect corresponds to a factor of $\sim2$ difference between $\log M/M_{\odot}\sim10.4$ and $\log M/M_{\odot}\sim11.4$ or $R_e\sim1$kpc and $R_e\sim10$kpc. 

\begin{figure}[t]
    \centering
    \includegraphics[width=0.45\textwidth]{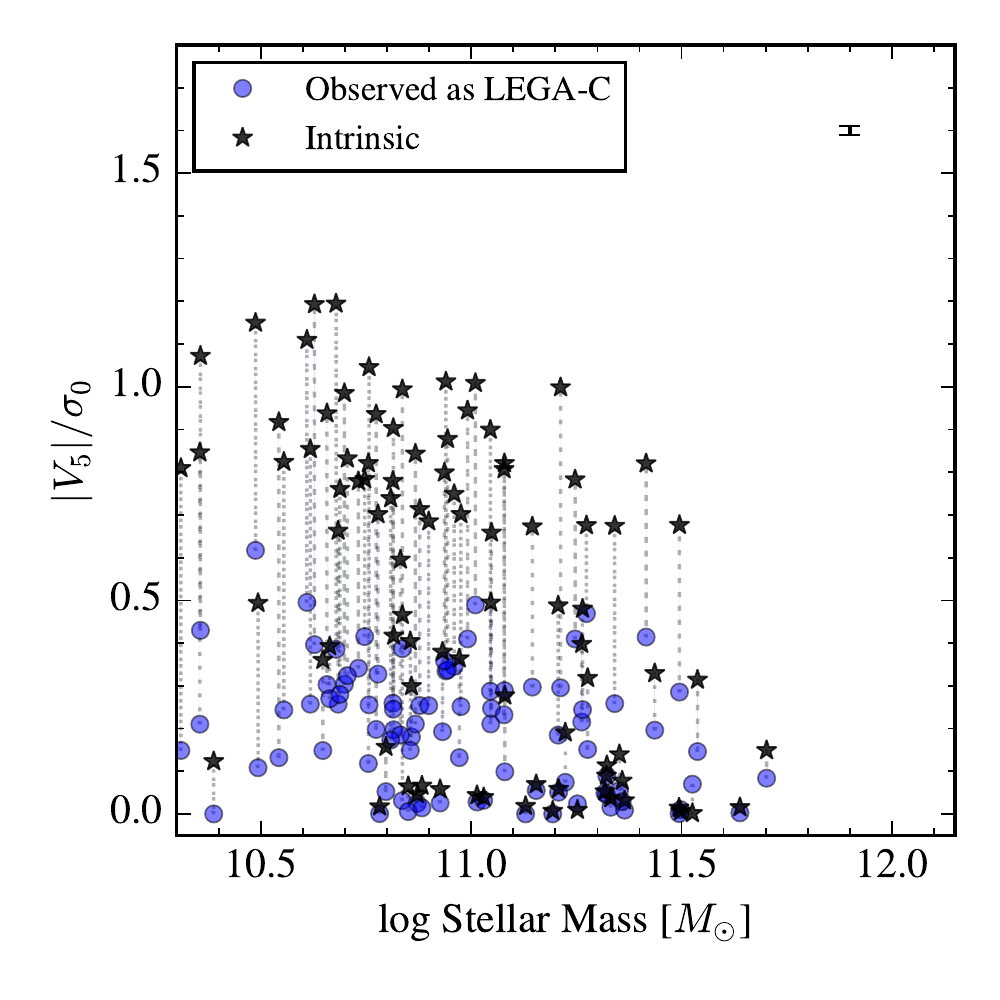}
    \caption{Rotational support ($|V_5|/\sigma_0$) versus stellar mass of CALIFA galaxies. The intrinsic values, as measured along the photometric position angle{, are} indicated by black stars. The average uncertainty in this measurement is indicated by the black errorbar in the upper right corner. The $|V_5|/\sigma_0$ for each galaxy with a misaligned slit and after convolution with a Moffat PSF ($\mathrm{FWHM}_{PSF}=7$kpc) is indicated by a blue circle, with measurements for each galaxy connected by gray dotted lines. Because the simulated PSF is significant relative to the physical extent of galaxies, the measured rotational support is strongly affected by the observational effects.}
    \label{fig:mstar_vsig_blur}
\end{figure}
\begin{figure*}[t]
    \centering
    \includegraphics[width=\textwidth]{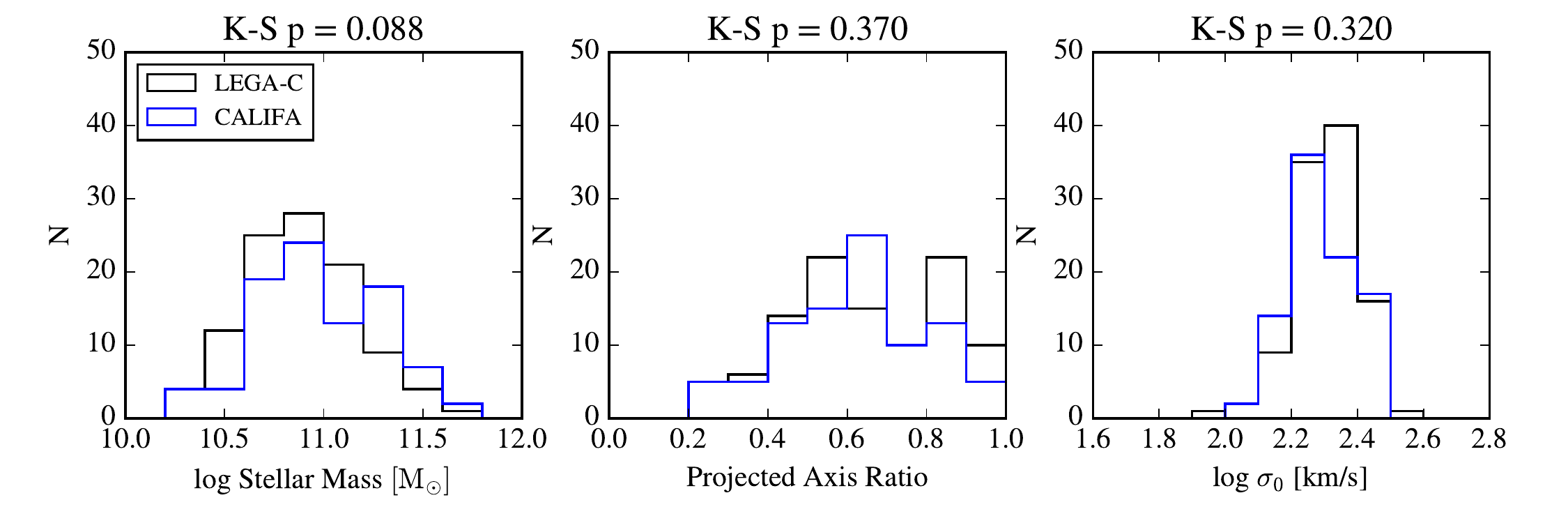}
    \caption{The distributions of LEGA-C (black) and CALIFA (blue) datasets in stellar mass (left panel), projected axis ratio (center panel), and velocity dispersions (right panel). The p-value of a two-sample K-S test is indicated at the top of each panel. Although the stellar mass distributions are not consistent at the 10\% level between the two samples, this property is sensitive to subtle differences in the modelling of the stellar populations between the two samples. The projected axis ratio and velocity dispersions are consistent with being drawn from the same distributions between the two surveys, with p=37\% and { p=32\%} respectively.}
    \label{fig:hists}
\end{figure*}

The overall effects of beam-smearing and misalignment are presented in Figure \ref{fig:mstar_vsig_blur}, which shows $|V_5|/\sigma_0$ versus stellar mass. Black stars indicate the intrinsic values along the photometric axis, connected by black dashed lines to blue circles from the simulations. The PSF preferentially decreases the observed rotation in lower mass galaxies. Furthermore, these observational effects lower the observed range in $|V_5|/\sigma_0$, thereby diminishing the dichotomy between slow and fast rotating galaxies. Although the intrinsic $|V_5|/\sigma_0$ measurements extend to much higher values ($>1.0$), all simulated $|V_5|/\sigma_0$ values are below $0.5$ and the correlation between stellar mass and rotational support is all but erased.

\subsection{Measuring redshift evolution}

Armed with the CALIFA sample of $z\sim0$ galaxies for which we have similar measurements of rotational support and have simulated the observational effects that are impacting the LEGA-C observations, we now move to assess the redshift evolution of the rotational support of quiescent galaxies. Before comparing the two samples, we would like to verify that they span similar regions of parameter space. In Figure \ref{fig:hists} we show the distributions of the CALIFA and LEGA-C quiescent samples in stellar mass (left panel), projected axis ratio (center panel), and central velocity dispersion (right panel).  In each case we perform a two-sample K-S test to evaluate whether the two samples are likely to be drawn from the same distributions. The K-S tests suggest that the samples are very well matched in projected axis ratio and velocity dispersion, while the stellar mass distributions are slightly different, with a p-value of 0.088, but not at a statistically significant (e.g. 3$\sigma$) level.  We emphasize that stellar masses are extremely sensitive to differences in modeling of the photometry and the stellar populations, whereas the other two properties ($b/a$ and $\sigma_0$) are measured reasonably consistently between the two samples and are generally less sensitive to systematics. In particular, we note that this difference also complicates comparisons between the two samples at fixed mass. We conclude that the CALIFA and LEGA-C samples are reasonably well-matched in axis ratio distribution and gravitational potentials to test redshift evolution of ${|V|/{\sigma}}$.

\begin{figure*}[t]
    \centering
    \includegraphics[width=0.99\textwidth]{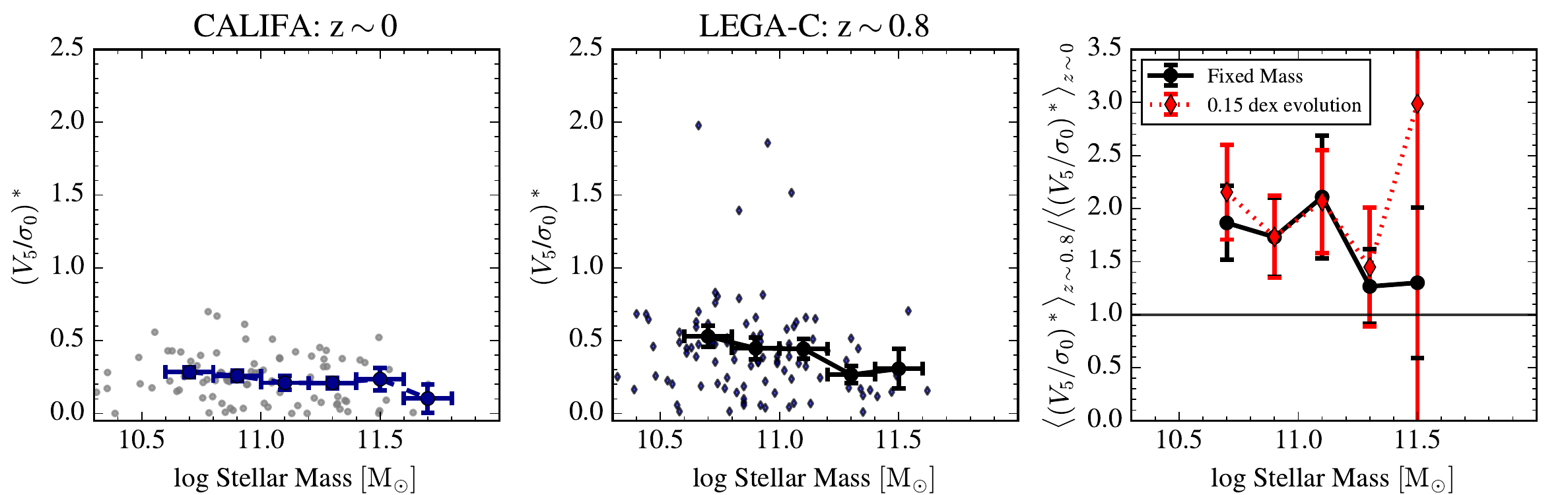}
    \caption{Rotational support ($(V_5/\sigma_0)^*$) versus stellar mass for the simulated CALIFA $z\sim0$ galaxies (left panel), LEGA-C galaxies at $z\sim0.8$ (center panel), and the ratio of the averages (right panel). At the highest mass end ($\log M_{\star}/M_{\odot} \gtrsim 11.25$) the rotational support is very similar, but at lower masses, the LEGA-C sample exhibits similar or slightly more rotational support than CALIFA galaxies at fixed mass { (black symbols and solid line in the right panel)}. However, when compared to more massive descendants (red symbols and dashed line{, assuming 0.15 dex evolution}) galaxies at $z\sim0.8$ exhibit $50-100\%$ higher rotational support than local galaxies.}
    \label{fig:lmass_vsig_comp}
\end{figure*}

Figure \ref{fig:lmass_vsig_comp} shows the rotational support versus stellar mass for the simulated ``observed'' CALIFA galaxies at $z\sim0$ in the left panel (gray symbols, blue dashed line), the LEGA-C sample at $z\sim0.8$ in the center (black diamonds and black solid line), with the running averages on the individual panels and together on the right panel. Uncertainties in the averages are calculated using jackknife resampling. For these comparisons we use $(V_5/\sigma_0)^*$ as a measure of rotational support to minimize scatter introduced by projection effects. 

Below $\log M_{\star}/M_{\odot} \sim 11.25$ galaxies at high redshift exhibit slightly ($\sim50\%$) more rotational support than those in the CALIFA sample. At the highest masses, the two samples are nearly consistent within the measurement uncertainties. However, the structural evolution of massive galaxies \citep[e.g.,][]{bezanson:09,hopkins:09cores,naab:09,dokkum:10} and evolution of the stellar mass function \citep[e.g.,][]{muzzin:13} implies that galaxies must grow in mass through cosmic time. Therefore, evolution at fixed mass is likely an underestimate in the dynamical evolution of individual galaxies. Empirically motivated work \citep[e.g.,][]{dokkum:13,leja:13,patel:13a} and theoretical studies \citep{behroozi:13,torrey:15,torrey:16} have estimated mass growth rates of ${\sim}0.15$ dex for massive LEGA-C-like galaxies since $z\sim1$. Accounting for this would imply stronger evolution than the comparison at fixed mass (red dashed line in Figure \ref{fig:lmass_vsig_comp}).   

\begin{figure*}[t]
    \centering
    \includegraphics[width=0.99\textwidth]{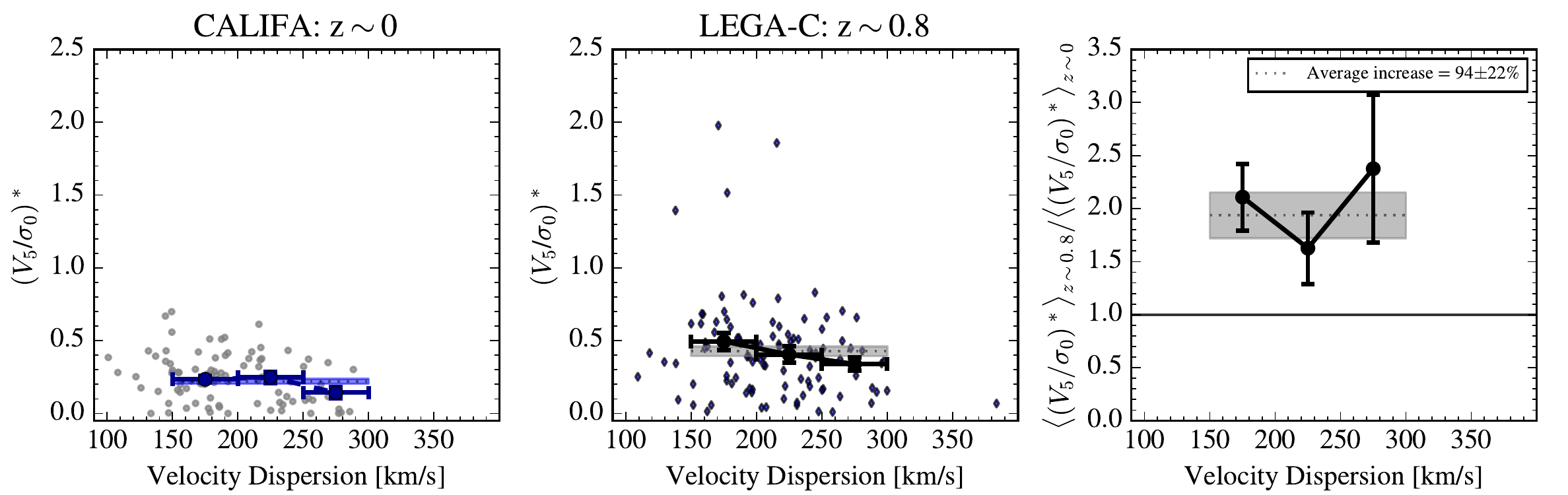}
    \caption{Rotational support ($(V_5/\sigma_0)*$) versus central velocity dispersion ($\sigma_0$) for the simulated CALIFA $z\sim0$ galaxies (left panel), LEGA-C galaxies at $z\sim0.8$ (center panel), and the ratio of the averages (right panel). { In each panel individual galaxies are indicated by small symbols, larger black circles indicate the running averages and jackknife uncertainties, and the horizontal bands indicate the average value and uncertainty evaluated between $150<\sigma<300$ km/s.}  At fixed central velocity dispersion, which is likely more stable than stellar mass, the higher redshift galaxies exhibit more rotational support than their local counterparts by a factor of $\sim1.5-2$.}
    \label{fig:sig_vsig_comp}
\end{figure*}

Another option is to compare at fixed central velocity dispersion, which may be a more stable property for an evolving galaxy \citep{bezanson:12,oser:11,sande:13,belli:14a,belli:14b}. We note that the effects of beam smearing can influence the measured central velocity dispersions, however from our simulations of the CALIFA stellar kinematics we expect this to be at most a few percent effect. Figure \ref{fig:sig_vsig_comp} follows the same conventions as Figure \ref{fig:lmass_vsig_comp}, but now compares rotational support to central velocity dispersion ($\sigma_0$). This also has the benefit of avoiding inconsistencies between samples in the SPS modeling used to estimate stellar mass. These panels indicate that for all galaxies at fixed velocity dispersion, rotational velocities are ${\sim}50-100\%$ higher at $z\sim0.8$ than in local quiescent galaxies, with an average ratio of ${1.94\pm0.22}$. Qualitatively, the observed evolution in rotational support at fixed velocity dispersion is robust to aperture and size evolution. In Appendix \ref{sect:V} we investigate the use of { two additional apertures. First, we adopt} an aperture that scales with the average effective radius, comparing rotation within 7.5 kpc for the CALIFA dataset at $z\sim0$ with $(V_5/\sigma_0)^*$ for the LEGA-C sample. Although the apparent evolution in rotational support is weaker than within a fixed aperture, with an average ratio of $\langle(V_{5}/\sigma_0)^*_{z{=}0.8}/(V_{7.5}/\sigma_0)^*_{z{=}0}\rangle = {1.41\pm0.16}$. This difference, significant at only the $\sim$95-percent level, is consistent with the results obtained using a more robust fixed aperture, and paint a similar picture that the degree of rotational support is higher than at the present day. { We also test the use of the maximum observed velocities ($V_{max}$) measured from each rotation curve, defined as the value of the best-fit arctangent function at the maximum extent, averaged symmetrically. This comparison yields an evolution of $\langle(V_{max}/\sigma_0)^*_{z{=}0.8}/(V_{max}/\sigma_0)^*_{z{=}0}\rangle = 1.76\pm0.22$, which is consistent with the evolution within 5 kpc at the ${\sim1}\sigma$ level.}

These results suggest a significant evolution in the rotational support of quiescent galaxies. At face value, this is consistent with results from the \citet{chang:13} study which found a decrease in the fraction of oblate rotators in massive ($10.8<M_{\star}/M_{\odot}<11.5$) quiescent galaxies from CANDELS/3DHST at $1<z<2.5$ to SDSS at $z\sim0.06$. Although the statistics in the study are somewhat small, they found a factor of ${\sim}2-4$ increase in the oblate fraction between SDSS and $0.6<z<0.8$ and $0.8<z<1.3$. However, this study also found no statistically significant evolution in the fraction of oblate rotators in the intermediate mass  ($10.5<M_{\star}/M_{\odot}<10.8$), where we observe an evolution in the $|V_5|/\sigma_0$ at fixed mass. 

We note here that there are subtle differences between the CALIFA and LEGA-C samples. Although we do not expect any to dominate the conclusions of this study, we mention them now for completeness. First, the distinction between quiescent and star-forming galaxies is defined differently for each sample: CALIFA uses spectroscopic criteria whereas for LEGA-C we use photometric colors. Secondly, although both surveys are initially magnitude limited, the CALIFA dataset is selected with an additional angular size selection to optimally utilize the IFU spectrograph. This latter selection will bias the CALIFA dataset against small galaxies and in particular will render the sample incomplete at the low mass end, however we note that at $9.7<\log M_{\star}/M_{\odot} < 11.4$ the overall CALIFA sample is representative in size \citep{walcher:14}, which safely includes the current sample. Furthermore, only 300 (${\sim}80\%$) of the CALIFA DR3 galaxies with V1200 grating data is included in the \citet{falconbarroso:17} stellar kinematics sample. Although the authors emphasize the representative redshifts, sizes, and absolute magnitudes of the resulting sample with respect to the full CALIFA dataset, the 75 galaxies that are eliminated due to poor quality stellar kinematic maps could introduce additional bias in the kinematic properties of quiescent galaxies. Finally, we have not attempted to match the LEGA-C and CALIFA samples in volume or environment or explicitly link individual progenitor and descendant galaxies. While we note that this could strengthen our conclusions about the redshift evolution of the dynamical structures of quiescent galaxies, this is beyond the scope of this paper.

\section{Discussion and Conclusions}\label{sect:discussion}

In this paper, we present the first results from spatially resolved stellar kinematics of a large sample of massive, quiescent galaxies at large lookback time, drawn from ESO's Public Spectroscopic LEGA-C Survey. As opposed to earlier work on smaller samples \citep{marel:07,moran:07,wel:08_rot} our sample is not selected on the basis of visual morphology, but rather by a lack of star formation, preventing a possible bias against disk-like, passive galaxies. The exceptional depth of the LEGA-C spectroscopic survey allows for spatially resolved kinematic modeling of the continuum beyond two effective radii of galaxies at $z\sim1$. 

We have demonstrated that galaxies at $z\sim0.8$ follow a similar trend of decreasing rotational support with increasing stellar mass as local early-type galaxies \citep[e.g.,][]{cappellari:07,emsellem:11}. But also like their local counterparts \citep[e.g.,][]{veale:17}, there exist examples of very massive fast-rotators in the LEGA-C sample. We find that ${\sim}90\%$ of very elongated galaxies, with projected axis ratios less than ${\sim}0.6$ exhibit significant rotation. The latter result adds credence to the empirical result that massive galaxies at high redshifts are more disk-like based on axis ratio distributions \citep[e.g.,][]{chang:13}. Conversely, the lack of a clear trend between rotation and S\'ersic index suggests that the concentration of a galaxy's light distribution is not a strong test of whether it is disk-like for quiescent galaxies. Furthermore, we emphasize that none of the properties (stellar mass, projected axis ratio, and S\'ersic index) explored in this work definitively predict the rotational support of an individual galaxy.

At fixed stellar velocity dispersion quiescent galaxies show $\sim 90\%$ more rotation on average within an aperture of radius 5 kpc at $z\sim 1$ than in the present-day universe. The most plausible interpretation is that such galaxies have lost angular momentum over the past 7 Gyr. Further interpretation of this observation in terms of evolution of individual galaxies is complicated by the fact that a significant number of galaxies cease star formation and join the quiescent population between $z\sim 1$ and the present. This `progenitor bias' \citep[e.g.,][]{franx:96,dokkum:00} forces us to consider that star-forming galaxies show a larger degree of rotational support than quiescent galaxies of the same mass or velocity dispersion. Hence, we can firmly rule-out the scenario that the cessation of star formation and the subsequent phase of evolution do not affect the dynamical structure: if that were the case, then we would see more rotational support amongst present-day quiescent galaxies compared to $z\sim 1$, instead of less. 

There are two (extreme) scenarios to explain the observed evolution in the $|V_5|/\sigma_0$ distribution at fixed mass and velocity dispersion (Figures \ref{fig:lmass_vsig_comp} and \ref{fig:sig_vsig_comp}). A first scenario is that galaxies drastically and suddenly lose their net angular momentum concurrently with the cessation of star formation. In this case, individual quiescent galaxies would not need lose angular momentum afterward to fit within the quiescent population. 
A second scenario is that angular momentum does not change in association with the cessation of star formation, and that quiescent galaxies gradually lose angular momentum through subsequent assembly, that is, dissipationless merging. In this scenario the fast-rotating quiescent galaxies would be younger than slow-rotating galaxies. Reality might well be a mixture of these scenarios. A follow-up study with the larger LEGA-C sample will explore these scenarios by comparing dynamical structure with stellar population ages.

At this time we can already surmise that the first scenario -- invoking rapid dynamical evolution -- appears unlikely to be the dominant mode of evolution. Only major mergers can accomplish sudden and drastic changes in dynamical structure \citep[e.g.,][and references therein]{naab:14} and few are seen among the star-forming population \citep[e.g.,][]{lotz:11,man:16}. This mode of transformation is not firmly ruled out however, as the timescales used to translate between pair fractions or disturbed morphologies and merger rates remain somewhat uncertain. Given that merging timescales are short, if all galaxies that were to become quiescent via mergers, the merger fraction implied is potentially close to the observed one \citep[e.g.,][]{bell:06,robaina:10}.  The second scenario -- gradual loss of angular momentum -- and the observed decline in the number density of very compact galaxies \citep{trujillo:09,taylordearth} provide mutual support for dissipationless growth of quiescent galaxies. 

Previous studies with morphologically selected galaxies have found contradictory results. \citet{wel:08_rot} found no evidence for an evolution in the rotational support of 25 $z\sim1$ elliptical and S0 galaxies using Jeans modeling to determine intrinsic rotational velocities and velocity dispersion profiles under the assumption that mass-follows-light and axisymmetric orbits. However, using similar analysis, \citet{marel:07} found an increase in rotation rates of cluster elliptical galaxies at $z\sim0.5$ at a confidence level of ${\sim}90\%$. Our analysis is consistent with the \citet{marel:07} study, however we note several key differences that prevent a direct quantitative comparison of our results to the previous work. First of all, we do not attempt to derive intrinsic, yet model-dependent, properties of the galaxies in the LEGA-C sample. Instead, we self-consistently simulated the observational effects of seeing, slit geometry, and binning on our low-redshift sample. One of the possible explanations cited for the discrepancies between the two previous studies is the different treatment of morphological classifications and potential misclassification of S0 galaxies in the \citet{marel:07} study. In contrast, in this study we do not distinguish amongst morphological classes of galaxies: all galaxies with quenched star-formation will fall into the high and low redshift samples, somewhat eliminating such progenitor biases. Of course one cannot avoid the bias introduced by excluding galaxies that are still forming stars.

\citet{belli:17} found an evolution in the dynamical to stellar mass ratios versus axis ratios of disky quiescent galaxies, as defined by their S\'ersic indices. They concluded that this difference implies an evolution in the average rotational support from $z\sim1.5-2.5$ until $z\sim0$, with the characteristic $|V|/\sigma$ decreasing from ${\sim}3$ to ${\sim}1.5$. This result is qualitatively consistent with our observed evolution given that the \citet{belli:17} sample is at higher redshifts than the LEGA-C sample. However, we note that we find S\'ersic index to be a very poor predictor of $V/\sigma$, especially at disk-like ($n<2.5$) values. 

Our result is consistent with predictions from simulations of isolated galaxy mergers \citep[e.g.,][]{wuyts:10}, the remnants of which have higher $V/\sigma$ at fixed ellipticities than local galaxies. Semi-analytic models constructed to explain the formation of fast and slow rotating early type galaxies in the \mbox{ATLAS3D} sample predict strong evolution in the number densities of fast and slow rotators with time: implying an increase by 0.7 dex in the number density of slow rotating galaxies and 0.2 dex for fast rotators since $z\sim1$ for massive galaxies $\log M_*/M_{\odot}> 11$ \citep{khochfar:11}.  Furthermore, a smooth evolution of rotational support is apparent within the Illustris cosmological simulation for massive galaxies that exhibit little rotation at $z\sim0$ \citep{genel:15}. In this and other simulations \citep[e.g.,][]{naab:14}, this evolution is due to a sequence of substantial minor merging. In this scenario, even rapidly rotating galaxies can evolve through time to slow-rotators, highlighting the importance of our approach of not excluding galaxies that would be morphologically classified as late types in our comparison. 

The current analysis falls short of deriving intrinsic dynamics for individual galaxies. Joint modeling of the spatially resolved kinematics and HST/ACS imaging including Jeans modeling, and assessment of the PSF size for each individual galaxy, to derive intrinsic properties would allow for a different direct comparison of the resolved kinematics of LEGA-C galaxies to local fast and slow rotating elliptical galaxies. This modeling is outside of the scope of the current paper, but is underway. Furthermore, this sample is only based on the first year LEGA-C data. Over the next few years, the full survey will be completed and the sample will increase by a factor of ${\sim}4$. Uncertainties in this kinematic modeling could be assessed by observing a subset of the current sample of galaxies with a perpendicular slit (East-West).

We have not addressed the question of whether rotational support depends on how recently a galaxy has quenched its star-formation. If rotation is diminished via minor merging, one might expect to see differences in the stellar ages or metallicities between slow and strongly rotating galaxies. Furthermore, we have not investigated evolution within the ${\sim} 2\,\mathrm{Gyrs}$ probed by the $0.6<z<1$ LEGA-C redshift range. Performing these tests while holding constant other properties that correlate with rotation would be difficult with the current sample of ${\sim}100$ LEGA-C galaxies. However, with the complete dataset we will test correlated trends in stellar age, rotational support, size, and stellar mass to test whether newer additions to the red sequence exhibit predicted differences from their older counterparts.

Ideally, one would like to observe the rotational support of quiescent galaxies as close to their epoch of transformation as possible. Below $\log M_{\star}/M_{\odot }\lesssim11$, where we expect galaxies to continue to grow and evolve below $z\sim1$, the LEGA-C dataset will probe stellar kinematics for star-forming progenitors and quiescent galaxies alike, in addition to any observable intermediate stages. For the most massive galaxies, we expect this to be at a much earlier epoch at $z\sim2-4$ from either stellar ages and colors \citep[e.g.,][]{kriek:08,whitaker:12a,mcdermid:15,glazebrook:17}. However at these redshifts continuum spectroscopy is extremely difficult, even with the latest generation ground-based Near-IR spectrographs. Spatially resolving the stellar continuum has only been possible for a few strongly lensed quiescent galaxies \citep[][Newman et al., in prep]{newman:15,toft:17}, unfortunately the low number density of massive quiescent galaxies will always render such targets extremely rare. The best hope for obtaining spatially resolved stellar kinematics in the near future is via deep spectroscopy with NIRSPEC on JWST or with adaptive-optics assisted observations on thirty-meter class telescopes.

\acknowledgments

RB would like to thank M. Cano-D\'iaz for providing electronic data tables from her paper and Guillermo Barro, Sirio Belli, Danilo Marchesini, Ryan Quadri, Nic Scott, and the many participants and organizers for entertaining and productive conversations at the the "Deconstructing Galaxies at Cosmic Noon: The Present and Future of Deep Spectroscopic Surveys at High Redshift" Lorentz Center Workshop. CP acknowledges support by an appointment to the NASA Postdoctoral Program at the Goddard Space Flight Center, administered by USRA through a contract with NASA. This research made use of Astropy, a community-developed core Python package for Astronomy \citep{astropy}. Based on observations collected at the European Organisation for Astronomical Research in the Southern Hemisphere under ESO programme 194.A-2005. This study uses data provided by the Calar Alto Legacy Integral Field Area (CALIFA) survey (http://califa.caha.es/). Based on observations collected at the Centro Astron\'omico Hispano Alem\'an (CAHA) at Calar Alto, operated jointly by the Max-Planck-Institut f\"{u}r Astronomie and the Instituto de Astrof\'isica de Andaluc\'ia (CSIC).

\appendix

\section{CALIFA 1D Rotation and Velocity Dispersion Profiles}\label{sect:califa_rot}

\begin{figure*}
    \centering
    \includegraphics[width=0.92\textwidth]{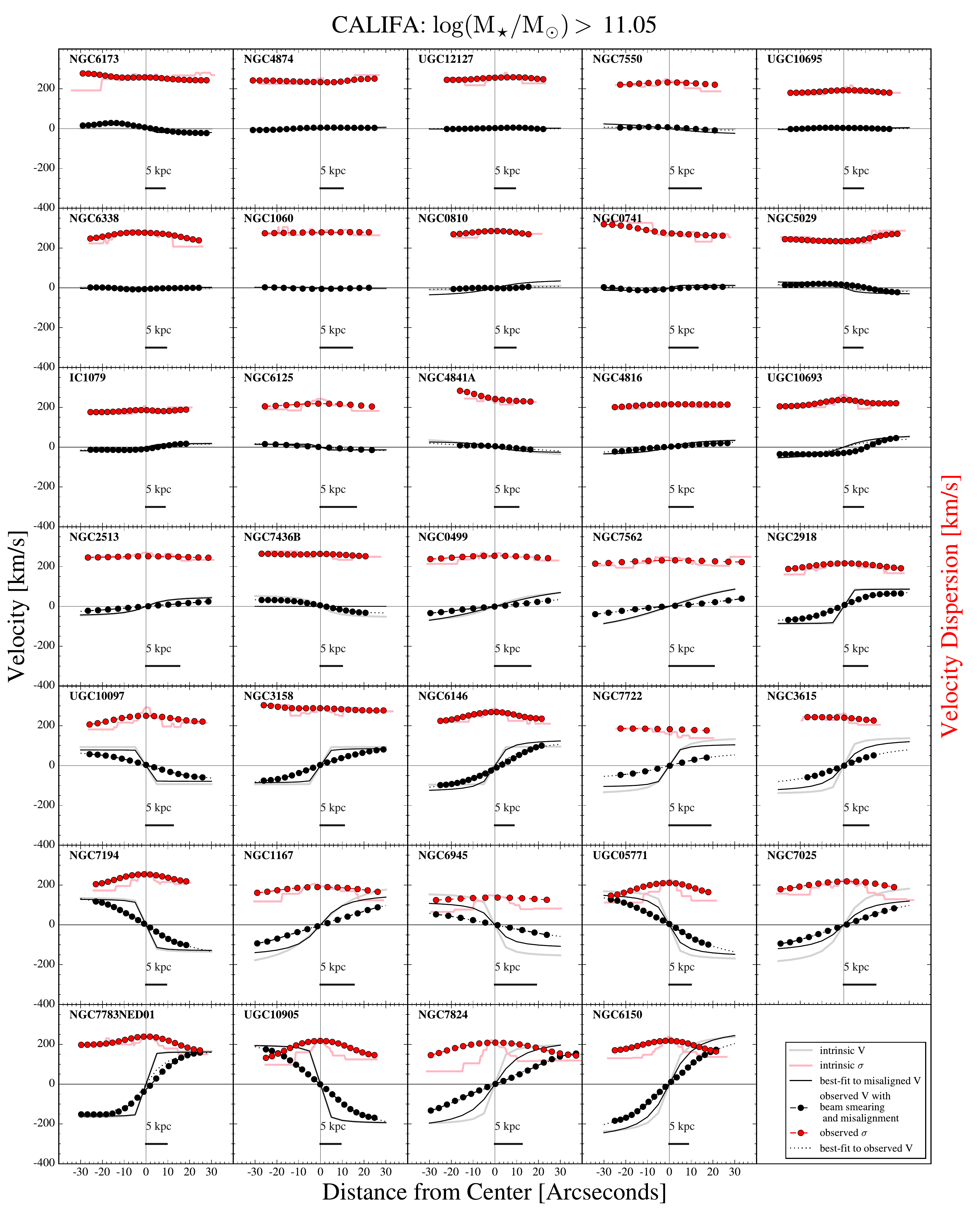}
    \vspace{-10pt}
    \caption{Stellar rotation curves (black) and velocity dispersion profiles (red) for the highest mass ($\log M_*/M_{\odot} > 11.4$) sample of quiescent massive galaxies in the CALIFA stellar kinematics sample, ordered by ascending velocity. Rotational velocity is measured from best-fit arctangent functions at a radius of 5kpc from the central pixel. Measured rotational velocities and velocity dispersions in Voronoi bins along the position angle are indicated by light gray and pink solid lines respectively{, which we refer to as the intrinsic values}. Best-fit arctangent rotation curve along { the N-S or E-W simulated LEGA-C misaligned position angle is shown as solid black line. The measured velocity and velocity dispersions within LEGA-C sized ``pixels'' and including the effects of slit misalignment and beam smearing is shown by black and red symbols and best-fit arc tangent function to this simulated rotation curve is shown by black dotted line.}}
    \label{fig:rot_califa}
\end{figure*}

\begin{figure*}
    \centering
    \addtocounter{figure}{-1}
    \includegraphics[width=0.92\textwidth]{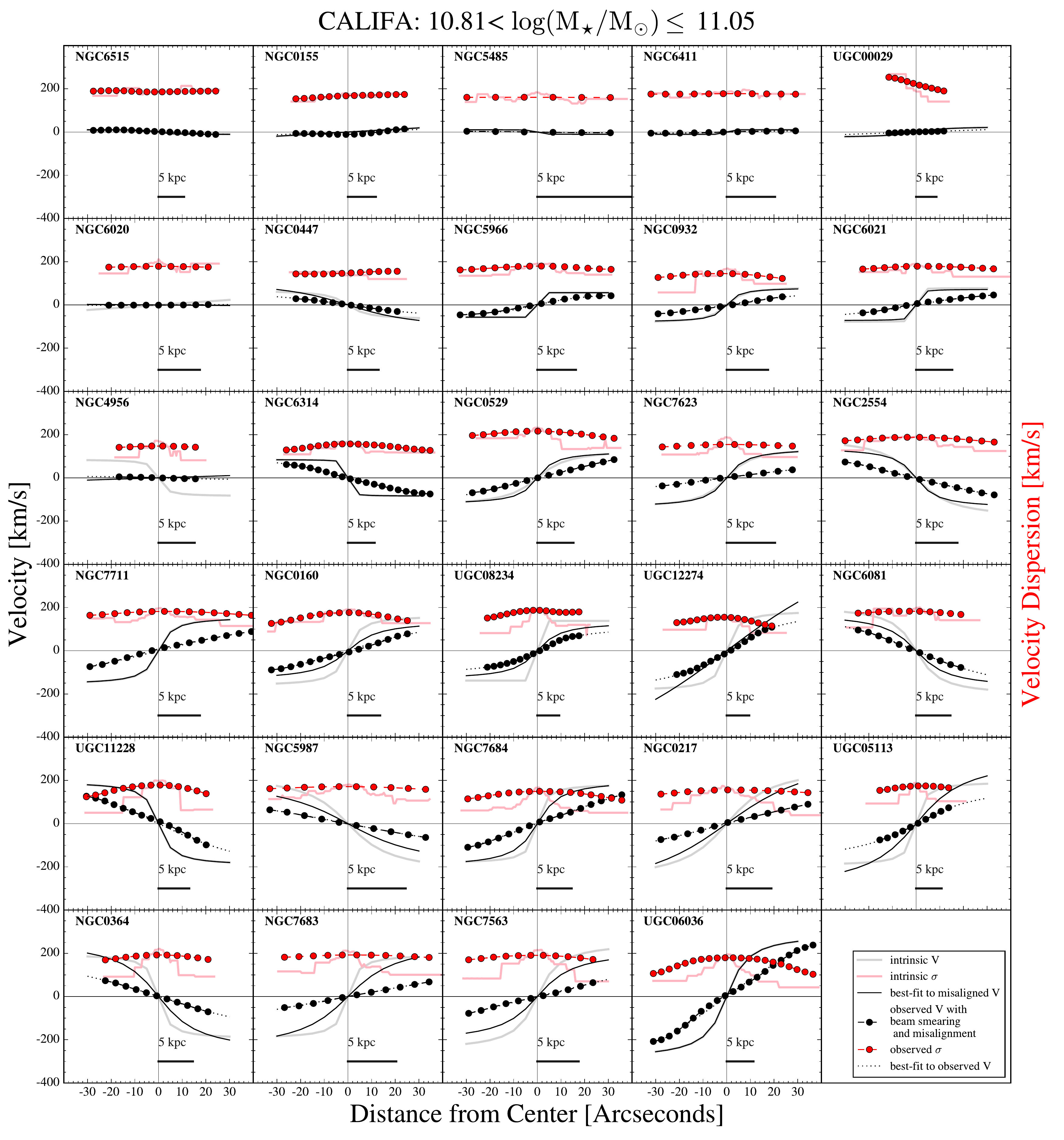}
    \caption{(Continued) Stellar rotation curves (black) and velocity dispersion profiles (red) for intermediate-mass quiescent massive galaxies in CALIFA, ordered by ascending velocity.}
\end{figure*}

\begin{figure*}
    \centering
    \addtocounter{figure}{-1}
    \includegraphics[width=0.92\textwidth]{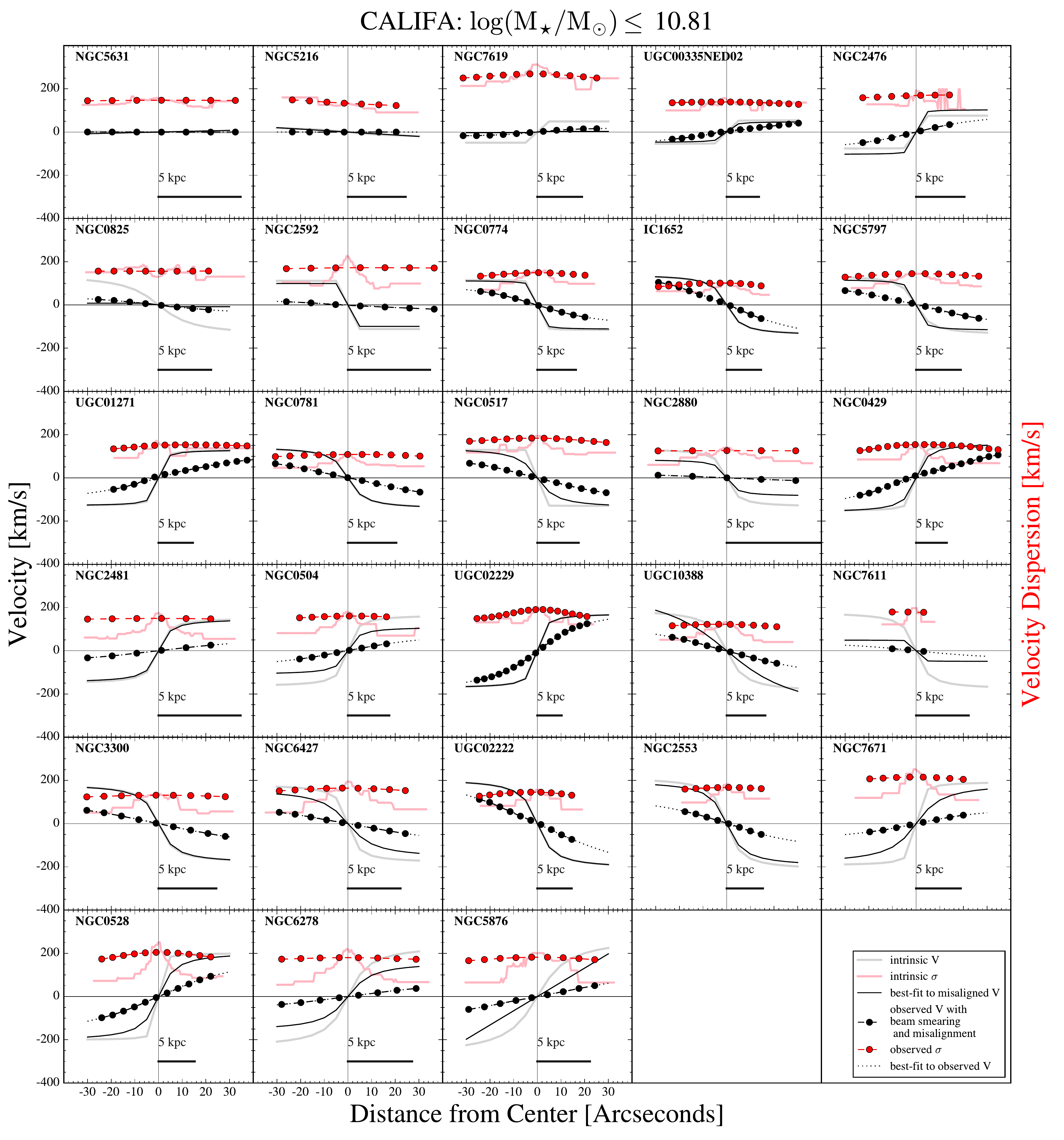}
    \caption{(Continued) Stellar rotation curves (black) and velocity dispersion profiles (red) for the lowest mass quiescent galaxies in CALIFA, ordered by ascending velocity.}
\end{figure*}

In this section we provide the one dimensional profiles derived from the CALIFA stellar kinematics data cubes \citep{falconbarroso:17}, as described in \S \ref{sect:califa}. All line-of-sight velocity and velocity dispersion profiles are included in Figure \ref{fig:rot_califa}, split into three pages in bins of descending mass and ordered by increasing rotational velocity at 5 kpc, $|V_5|$. The 1D profiles are extracted both along the position angle derived from the SDSS imaging in the galaxy outskirts \citep{walcher:14} and along the closer of the horizontal or vertical axes to approximate the LEGA-C North-South slit positions and position angle threshold for the analysis in this paper. The profiles are included for all 91 retired CALIFA galaxies in Figure \ref{fig:rot_califa}.  The quantities determined along the misaligned slit are indicated by black (velocity) and red (velocity dispersion) symbols. The rotation curve is fit with an arctangent function and this fit is indicated by a black, solid line. The velocity dispersion profiles measured along the position angle are shown as pink lines and the best-fit arctangent function fit to the rotation curve at the position angle are included as a gray solid line. The dependence of the measured velocity dispersion profiles on position angle is negligible and central velocity dispersions differ by ${\sim}1\%$ on average. The { velocity profiles exhibit a stronger dependence on position angle, but the effect is small, comprising a median decrease in $|V_5|/\sigma_0$ of 8\%}. We emphasize that this does not account for any possible misalignment between the kinematic and photometric axes, which can be misaligned by as much as 50\% \citep{emsellem:07}. However we note that this is not a dominant effect for this study, as the substantial effects of beam smearing due to the significant size of the PSF relative to the galaxy sizes in the LEGA-C sample dwarf the effects of up to 45 degrees of misalignment in this exercise. The profiles are also shown after convolution and luminosity-weighted extraction within a 7.5 kpc - wide slit as black (velocity) and red (velocity dispersion) dashed lines. A line in included in the lower left of each panel indicating the fixed physical scale of 5 kpc at which the velocity $V_5$ is measured.

\section{On the Choice of Velocity Aperture}\label{sect:V}

In this paper we adopt a measure of rotation within a fixed physical aperture, both for studying the properties of the LEGA-C sample and for comparison with the CALIFA dataset. This has two primary advantages. First, measuring velocity within a fixed aperture of 5kpc means that the bias on the measured rotational velocity introduced by beam smearing will be roughly the same for all galaxies in the sample; the rotational velocity within an aperture that scales with the effective radius of a galaxy would be impacted differently for large and small galaxies. Second, we expect the sizes of massive galaxies to evolve through cosmic time \citep[e.g.,][]{wel:14}. We believe that that evolution is largely inside-out growth driven by minor-merging \citep[e.g.][]{bezanson:09,hopkins:09cores,dokkum:10}, especially at $z\leq1$ where this study is focused \citep{newman:12}. In this framework, we would expect galaxies to grow in size, and mass by building up a more diffuse envelope around a dense central core. By focusing on the rotational support within a \emph{fixed} physical aperture, we probe the physical evolution of the same region of the galaxy, minimizing the additional confusion of whether the rotational support is physically evolving or rather whether it is the effect of using redshift-evolving aperture.

\begin{figure}
    \centering
    \includegraphics[width=0.45\textwidth]{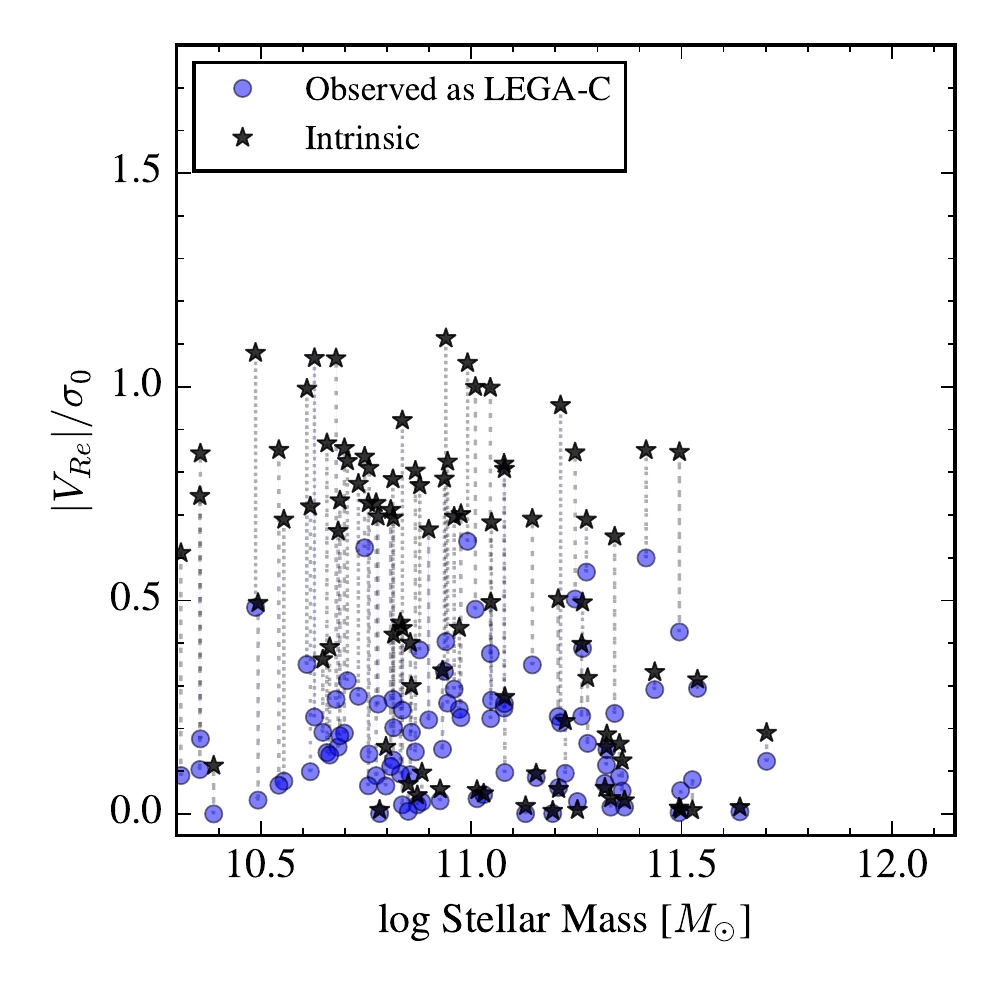}
    \includegraphics[width=0.45\textwidth]{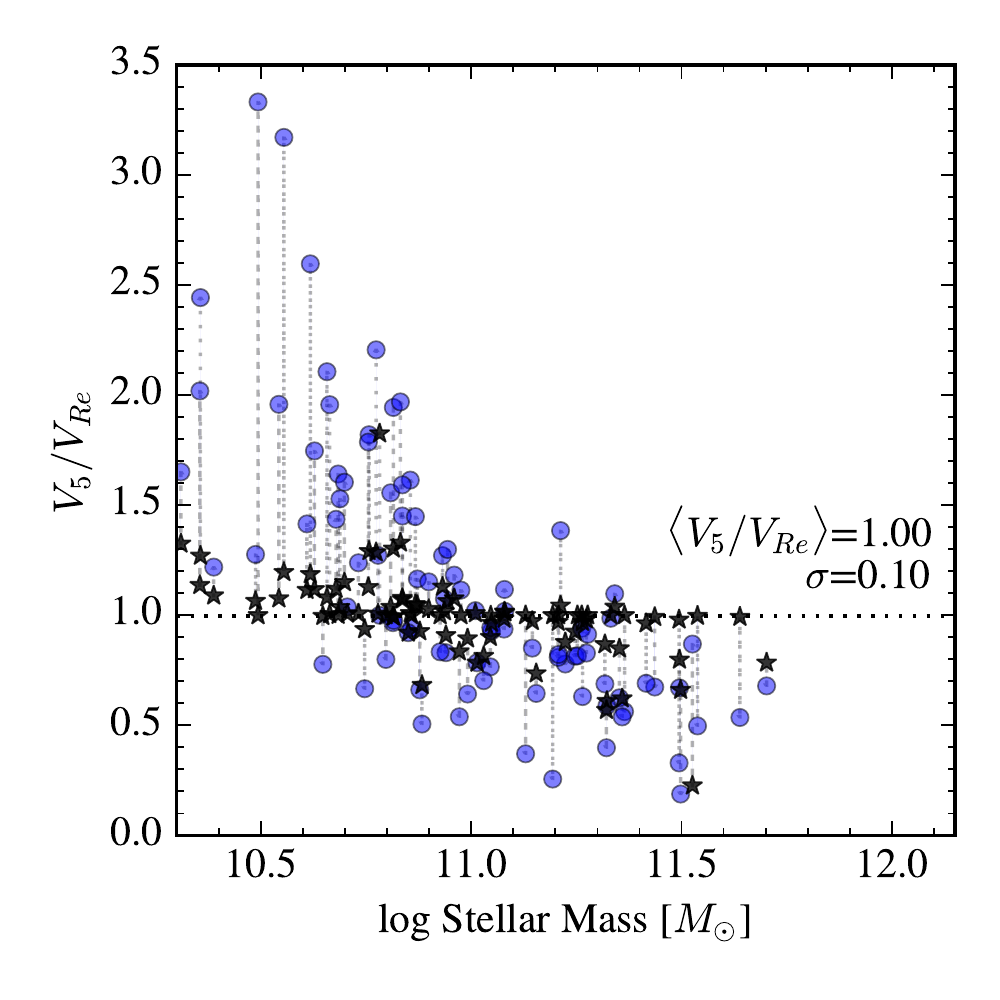}
    \caption{Measured velocity within an effective radius in CALIFA galaxies measured directly from the stellar kinematic maps (black) and from those blurred with the 3~kpc PSF (blue). As in Figure \ref{fig:mstar_vsig_blur}, this definition of reveals that the rotational support within the CALIFA dataset is a strong function of stellar mass. However, unlike a velocity defined within a fixed aperture, the effect of beam smearing on the measured velocity within an effective radius is significantly stronger for less massive, and therefore more compact, galaxies.}
    \label{fig:mstar_vsig_blur_re}
\end{figure}

However, in this appendix, we investigate the effects of defining the velocity at the effective radius ($V_{Re}$) of each galaxy. To demonstrate the impact of this, we begin by recreating the $|V|/\sigma$ versus stellar mass relation for the CALIFA galaxies in Figure \ref{fig:mstar_vsig_blur_re} using the velocity calculated within an effective radius ($V_{Re}$). The left panel shows the measured $|V_{Re}|/\sigma$ before and after convolution, similar to Figure \ref{fig:mstar_vsig_blur}.  Although the intrinsic points (black stars) show a very similar inverse correlation with stellar mass as with the velocity measured within 5~kpc, after convolving with a 3~kpc PSF the measured rotational support for low mass galaxies, for which half-light radii are smaller than 5~kpc, are dramatically diminished. We expect the effect to be even stronger at $z\sim0.8$, where LEGA-C galaxies are even more compact at fixed mass. The right panel of Figure \ref{fig:mstar_vsig_blur_re} shows the ratio of the two velocity measures versus stellar mass from the intrinsic (black stars) and blurred kinematic maps (blue circles).  The excellent agreement between the intrinsic $V_5/V_{Re}$, with an average ratio of 1.0 and a scatter of $\sim0.1$, suggests that either measure can be used to reliably assess the degree of projected rotation in the CALIFA dataset. However, after PSF convolution, the two measures diverge dramatically, with a very clear effect as a function of stellar mass. 

\begin{figure}
    \centering
    \includegraphics[width=\textwidth]{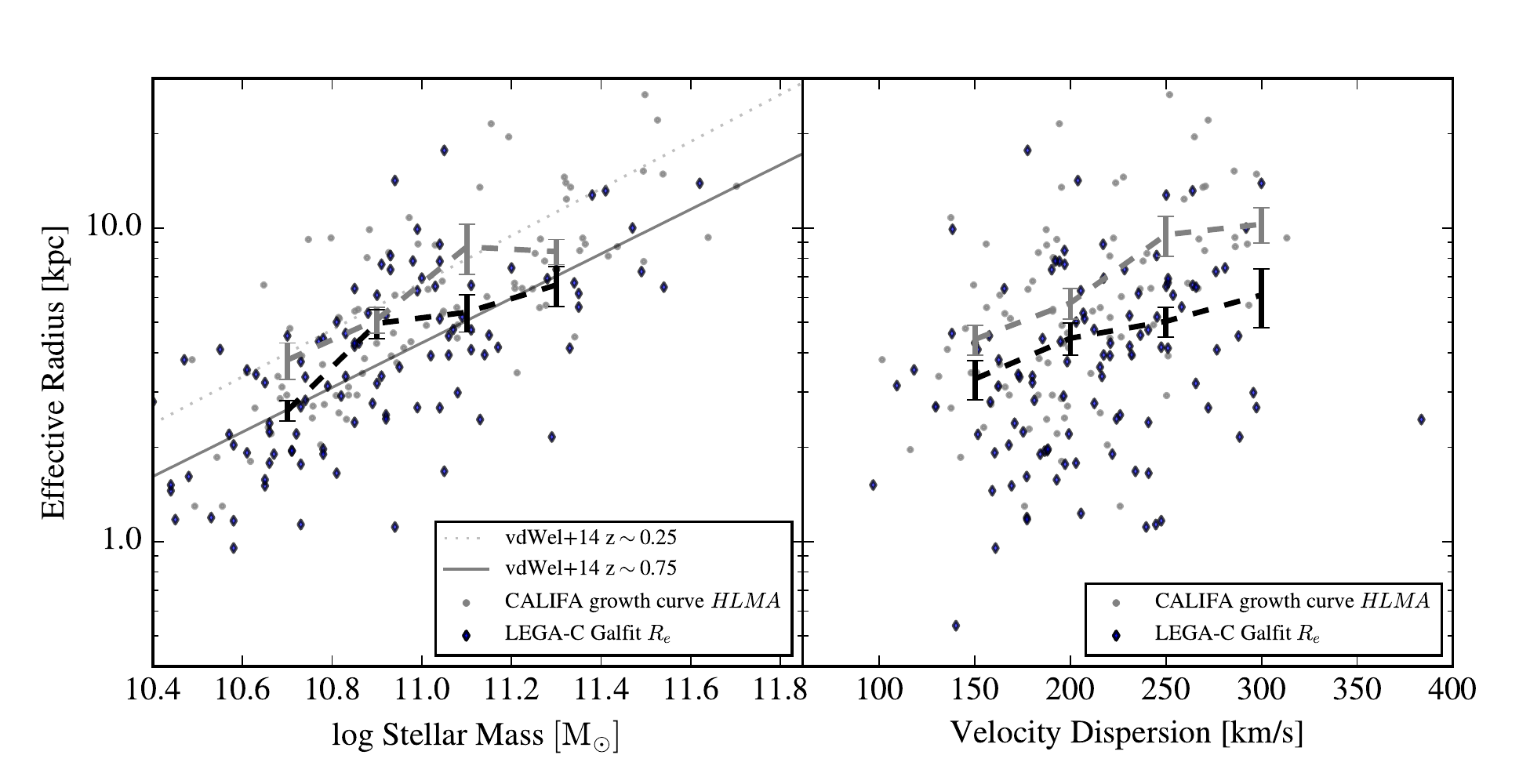}
    \caption{Size versus stellar mass (left panel) and velocity dispersion (right panel) for CALIFA (half-light major axis radius, black) and LEGA-C (S\'ersic half-light radius, gray) datasets. The running mean and scatter are indicated by dashed gray and black lines and the measured size-mass trends from \citet{wel:14} at $z\sim0.25$ and $z\sim0.75$ are indicated by dotted and solid lines respectively. At fixed mass and velocity dispersion CALIFA galaxies are more extended than those in the LEGA-C survey, confirming the expected trend of size evolution in the population of massive, quiescent galaxies.}
    \label{fig:size_mass_sigma}
\end{figure}

The second issue we wish to address in this appendix is our use of a fixed physical aperture and its impact on the measured evolution of rotational support. Figure \ref{fig:size_mass_sigma} shows the size versus stellar mass (left panel) and velocity dispersion (right panel) relations for the LEGA-C and CALIFA samples. In this figure, size corresponds to the semi-major axis of a best-fitting S\'ersic model for the LEGA-C sample and for CALIFA is the half-light-major axis derived using growth curve analysis \citep{walcher:14}.  These two methodologies can yield biased size measurements, \citet{mendezabreu:17} found significant offsets between single S\'ersic fits and HLMA for CALIFA galaxies, which can differ by up to a factor of $\sim2.5$, a scatter of $\sim20\%$, and a bias towards larger S\'ersic $Re$ than growth curve-derived HLMA. Therefore the relative evolution could be even stronger than suggested by this comparison. Furthermore, this measurement discrepancy is an additional factor in avoiding the use of a velocity aperture that scales with galaxy size. Given that we see clear evolution in galaxy sizes, we emphasize that only by measuring within a fixed physical aperture can we probe intrinsic evolution in the stellar orbits and rotational support. This necessarily implies that we are measuring the rotational support within a different fraction of galaxies at the two epochs. However, the combination of the observational effects inherent in our measurements and the observed size evolution between $z\sim0$ and $z\sim0.8$ leads us to conclude that measuring the velocity within a fixed physical aperture is the best course of action, settling on 5~kpc which roughly equals the extent of the least extended rotation curves in both surveys.

\begin{figure*}[t]
    \centering
    \includegraphics[width=0.99\textwidth]{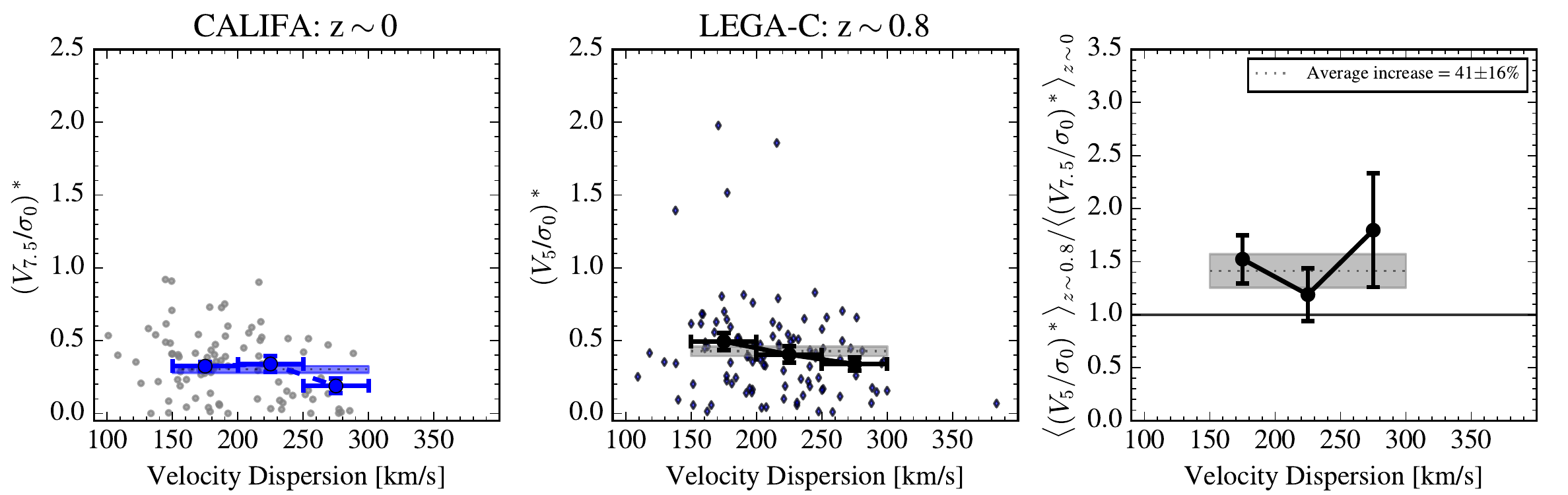}
    \caption{Rotational support ($(V/\sigma_0)*$) versus central velocity dispersion ($\sigma_0$) within an aperture that scales with increasing radius between the two samples. The left panel includes $|V_{7.5}|/\sigma_0$ for the simulated CALIFA $z\sim0$ galaxies and the center panel includes LEGA-C galaxies at $z\sim0.8$, for which velocities are measured within 5kpc. The right panel shows the ratio of the averages in small running bins (indicated by black points with errorbars) and the overall (for $150<\sigma/\mathrm{km/s}<300$) given by the gray band. Even within an aperture that scales with average effective radius, $|V|/\sigma$ is higher by ${41\pm16\%}$ at $z\sim0.8$.}
    \label{fig:sig_vsig_comp_r_evolve}
\end{figure*}

From Figure \ref{fig:size_mass_sigma} it is clear that a 5kpc aperture corresponds to a larger fraction of the LEGA-C galaxies than for the galaxies in the CALIFA sample. The average size of galaxies in the CALIFA sample is 1.5 times larger than in the LEGA-C sample. Given that many of the CALIFA rotation curves are still rising at 5kpc, at least some fraction of the observed discrepancy in rotational support between the two samples could be driven by size evolution. To investigate this effect, we scale the velocity aperture for the CALIFA dataset to 7.5kpc and compare $V/\sigma_0$ at fixed velocity dispersion to those measured for the LEGA-C sample within 5kpc. The results of this test are shown in Figure \ref{fig:sig_vsig_comp_r_evolve}. The left panel of this figure shows $|V_{7.5}|/\sigma_0$ versus velocity dispersion for CALIFA galaxies and the center panel of this figure shows $|V_{5}|/\sigma_0$ for LEGA-C galaxies. In each panel the running averages are indicated by solid blue and black lines with errorbars and the overall average and measurement uncertainty is included as a gray dotted line and horizontal band. All uncertainties in the averages are calculated via jackknife resampling. As expected from the rotation curves in Figure \ref{fig:rot_califa}, the $|V|/\sigma_0$ values measured within a larger physical aperture are higher on average than within 5kpc, however the LEGA-C sample still exhibits slightly more rotational support. The right panel shows the ratio of the running and overall averages, indicating that even when the velocity aperture is scaled to reflect the size evolution between the two epochs, we detect a ${\sim41\pm16\%}$ decrease in the rotational support of quiescent galaxies since $z\sim 1$. This implied evolution is more subtle than what we measure within a fixed physical aperture, but is still statistically significant. 

{ Finally, we investigate the use of the maximum measured velocity ($V_{max}$), defined as the value of the best-fitting arctangent function at the maximum extent of the measured rotation, averaged between the North and South directions.  In many cases, this might provide the best estimate of the intrinsic maximum rotational velocity. In Figure \ref{fig:sig_vsig_comp_max} we show the comparison between the rotational support measured at the maximum physical extent ($(V_{max}/\sigma_0)*$) versus velocity dispersion in the CALIFA and LEGA-C samples, as presented in Figures \ref{fig:sig_vsig_comp} and \ref{fig:sig_vsig_comp_r_evolve}. The maximum measured rotational support for CALIFA galaxies is shown in the left panel and for the LEGA-C sample in the center panel. Following the symbols and plotting conventions in previous figures, solid symbols indicate individual galaxies, large symbols indicate the average values in three velocity bins and the bands indicate the average $(V_{max}/\sigma_0)*$ between $150 < \sigma < 300$ in each sample. The right panel includes the ratio of the averages in three velocity dispersion bins (black error bars) and over the full range (gray band).  The rotational support ($(V_{max}/\sigma_0)*$) measured in this manner spans parameter space differently than defined at 5 kpc; values are higher on average at the maximum extent of the rotation curves than at 5 kpc. This is not surprising as most rotation curves in both samples do not flatten out (at LEGA-C seeing) and generally extend beyond 5 kpc.  However, the qualitative comparison between the CALIFA and LEGA-C rotational support yields a similar result with this $V_{max}$ definition as with $V_5$ (Figure \ref{fig:sig_vsig_comp}): galaxies in the LEGA-C sample exhibit $76\pm22\%$ higher average rotational support, measured by $(V_{max}/\sigma_0)*$, which is $\sim1\sigma$ below the comparison at a fixed 5 kpc aperture. We note that this aperture is less consistent within an individual sample (e.g. as a function of stellar mass, effective radius, or S\'ersic index) or between the two datasets, as the depths of the surveys are not perfectly matched. Therefore, although this may well come closer to the intrinsic values of $V_{max}$ for each individual galaxy, we rely primarily upon the 5 kpc aperture for the comparison in the main text of the paper. The qualitatively similar behavior for velocities measured within a variety of apertures
(5 kpc, evolving apertures to correct for galaxy size evolution, and at the maximum extend probed by the current data) lends to our confidence in the implied evolution in rotational support of quiescent galaxies since $z\sim1$.

\begin{figure*}[t]
    \centering
    \includegraphics[width=0.99\textwidth]{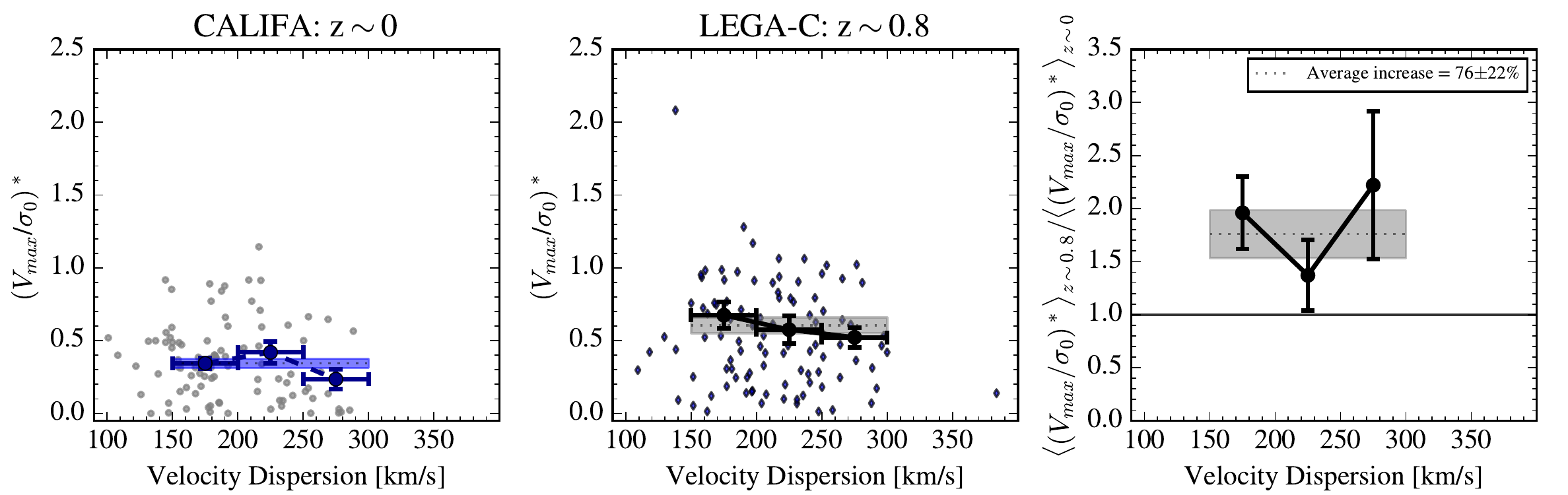}
    \caption{ Rotational support at the maximum radial extent ($(V_{max}/\sigma_0)*$) versus central velocity dispersion ($\sigma_0$). The left panel includes $|V_{max}|/\sigma_0$ for the simulated CALIFA $z\sim0$ galaxies and in the center panel for LEGA-C galaxies at $z\sim0.8$. The right panel shows the ratio of the averages in small running bins (indicated by black points with errorbars) and the overall (for $150<\sigma/\mathrm{km/s}<300$) given by the gray band. Within this maximum radius, $|V|/\sigma$ is higher by $76\pm22\%$ at $z\sim0.8$, which implies slightly less dramatic, but still significant rotation than for velocities defined within 5 kpc.}
    \label{fig:sig_vsig_comp_max}
\end{figure*}
}

\end{document}